\documentclass[preprint,12pt,authoryear]{elsarticle}

\usepackage{amssymb}
\usepackage{amsmath}
\usepackage{subcaption}
\usepackage[utf8]{inputenc}
\usepackage[T1]{fontenc}
\journal{International Journal of Plasticity}

\begin{document}

\begin{frontmatter}

%% Title, authors and addresses

%% use the tnoteref command within \title for footnotes;
%% use the tnotetext command for theassociated footnote;
%% use the fnref command within \author or \address for footnotes;
%% use the fntext command for theassociated footnote;
%% use the corref command within \author for corresponding author footnotes;
%% use the cortext command for theassociated footnote;
%% use the ead command for the email address,
%% and the form \ead[url] for the home page:
%% \title{Title\tnoteref{label1}}
%% \tnotetext[label1]{}
%% \author{Name\corref{cor1}\fnref{label2}}
%% \ead{email address}
%% \ead[url]{home page}
%% \fntext[label2]{}
%% \cortext[cor1]{}
%% \address{Address\fnref{label3}}
%% \fntext[label3]{}

\title{Strain localization and dynamic recrystallization in polycrystalline metals: thermodynamic theory and simulation framework}

%% use optional labels to link authors explicitly to addresses:
%% \author[label1,label2]{}
%% \address[label1]{}
%% \address[label2]{}

%\author{Charles K. C. Lieou, Hashem M. Mourad, and Curt A. Bronkhorst\corref{cor1}}
%\cortext[cor1]{Corresponding author}
%\ead{cabronk@lanl.gov}
%\address{Theoretical Division, Los Alamos National Laboratory, Los Alamos, NM 87545, USA}

\author{Charles K. C. Lieou}
%\address{Theoretical Division, Los Alamos National Laboratory, Los Alamos, NM 87545, USA}
\author{Hashem M. Mourad}
%\address{Theoretical Division, Los Alamos National Laboratory, Los Alamos, NM 87545, USA}
\author{Curt A. Bronkhorst\corref{cor1}}
\cortext[cor1]{Corresponding author}
\ead{cabronk@lanl.gov}
\address{Theoretical Division, Los Alamos National Laboratory, Los Alamos, NM 87545, USA\fnref{laur}}
\fntext[laur]{LA-UR release number: LA-UR-18-27644}

\begin{abstract}
%% Text of abstract
We describe a theoretical and computational framework for adiabatic shear banding (ASB) and dynamic recrystallization (DRX) in polycrystalline materials. The Langer-Bouchbinder-Lookman (LBL) thermodynamic theory of polycrystalline plasticity, which we recently reformulated to describe DRX via the inclusion of the grain boundary density or the grain size as an internal state variable, provides a convenient and self-consistent way to represent the viscoplastic and thermal behavior of the material, with minimal ad-hoc assumptions regarding the initiation of yielding or onset of shear banding. We implement the LBL-DRX theory in conjunction with a finite-element computational framework. Favorable comparison to experimental measurements on a top-hat AISI 316L stainless steel sample compressed with a split-Hopkinson pressure bar suggests the accuracy and usefulness of the LBL-DRX framework, and demonstrates the crucial role of DRX in strain localization.
\end{abstract}

\begin{keyword}
%% keywords here, in the form: keyword \sep keyword
Constitutive behavior \sep Dynamic recrystallization \sep Shear banding \sep Steel \sep Finite-element simulation \sep Taylor-Quinney coefficient
%% PACS codes here, in the form: \PACS code \sep code

%% MSC codes here, in the form: \MSC code \sep code
%% or \MSC[2008] code \sep code (2000 is the default)

\end{keyword}

\end{frontmatter}

%% \linenumbers

%% main text
\section{Introduction}
\label{sec:1}

Under severe loading conditions, adiabatic shear bands (ASBs) often develop in ductile metallic materials. The concentration of plastic deformation into narrow bands of the material, often preceding ductile fracture \citep[e.g.,][]{sabnis_2012,rousselier_2015,arriaga_2017}, has obvious implications for many industrial and defense applications such as metal-forming processes, shock absorption, and structural engineering. The need for an accurate representation of adiabatic shear bands, a thorough understanding of the physics of shear localization, and a predictive description of the formation and growth of shear bands has triggered a large amount of theoretical, numerical, and experimental research; excellent overviews are provided by the treatises of \citet{wright_2002} and \citet{dodd_2012}.

Numerical representation of ASBs presents a major challenge because of the short length and time scales involved. Mesh sensitivity issues naturally arise from direct finite-element implementations which impart a length scale equal to the element size in the shear band. Recent developments have made substantial progress in addressing these issues, through embedding the shear-band width into the problem by non-local techniques~\citep{anand_2012,ahad_2014,abed_2005,voyiadjis_2004,abed_2007,voyiadjis_2005,voyiadjis_2007,voyiadjis_2013}. A recent effort by two of us and collaborators \citep{mourad_2017,jin_2018} sought to eliminate mesh dependency through the sub-grid method, which permits the nucleation of shear bands narrower than the mesh size, effectively circumventing mesh dependency. 

The physical mechanisms underlying adiabatic shear localization pose a profound challenge very distinct from that on the numerical front; advances in numerical methods alone do not provide an understanding of ASB mechanisms, or a predictive description of the shear-banding process. There is growing evidence in the literature that dynamic recrystallization (DRX) -- the process by which fine, nano-sized grains with few or no dislocations form in the ASB -- provides an additional softening mechanism and supplements the role of thermal softening in shear band initiation. DRX has been observed in conjunction with adiabatic shear localization in a broad range of metals and alloys, including titanium and titanium alloys \citep[e.g.,][]{rittel_2008,osovski_2012,osovski_2013,li_2017}, magnesium \citep{rittel_2006}, copper \citep{rittel_2002}, and steel \citep[e.g.,][]{meyers_2000,meyers_2001,meyers_2003}.

Numerous authors have developed hypotheses for the underlying cause of dynamic recrystallization. \citet{brown_2012} attributes DRX to the diffusive motion of grain boundaries for relatively slow loading rates where diffusional time scales are relevant. For conditions of loading considered here, subgrain rotation and the lineup of dislocations forming new grain boundaries have been invoked as a contributing mechanism for the occurrence of DRX \citep{hines_1997,hines_1998,meyers_2000,meyers_2003,li_2017}. Others \citep[e.g.,]{popova_2015} have argued that DRX is probabilistic in nature and is driven by a local mismatch of the dislocation density. In light of these hypotheses, varied efforts have been made to develop theoretical descriptions of the DRX process. These include phenomenological models \citep[e.g.,][]{galindo-nava_2014,galindo-nava_2015,mourad_2017}, internal state variable theories \citep[e.g.,][]{brown_2012,puchi-cabrera_2018,sun_2018}, cellular automaton models \citep[e.g.,][]{popova_2015}, phase-field descriptions \citep[e.g.,][]{takaki_2008,takaki_2009}, or a hybrid of several of these such as a combination of phase-field modeling and crystal plasticity \citep[e.g,][]{zhao_2016,zhao_2018}.

Important physical ingredients seem to be inadequately included in many of the existing models in the literature. Firstly, many of these theories do not account for energy balance, a crucial physical ingredient in a deforming material because of the input work of deformation and the thermal effects arising from the input work. Some of these theories (e.g., the MTS model used in \citep{mourad_2017}) partition the stress into components accounting for different physical mechanisms such as the different types of barriers to dislocation motion. Because energy is conserved while there is no analogous conservation law for the stress, the input energy is the only quantity that can be additively partitioned into components stored or dissipated via different mechanisms without invoking additional implicit assumptions. Secondly, conventional theories employ flow rules that are largely phenomenological; examples include power-law fits of the type $\sigma \propto \dot{\epsilon}^{\delta}$ between the stress $\sigma$ and the strain rate $\dot{\epsilon}$, or between the strain rate and some internal slip resistance. These are based on extensive observations and a search for quantitative trends, and are good mathematical approximations; yet they shed no light on the underlying physical principles for such behavior, and may not apply with greater generality to other materials or loading rate regimes. Finally, while many of these models rightfully adopt a statistical view of dislocations, the assumed evolution of the dislocation density may be problematic on physical grounds. For example, the Kocks-Mecking equation \citep{kocks_1966,mecking_1981} for the temporal evolution of the dislocation density $\rho$, employed in many of the references \citep[e.g.,][]{takaki_2008,takaki_2009,zhao_2016,zhao_2018}, is of the form
\begin{equation}\label{eq:kocks}
 \dfrac{d \rho}{d \epsilon} = k_1 \sqrt{\rho} - k_2 \rho ,
\end{equation}
where $k_1$ and $k_2$ are parameters, and $\epsilon$ is the total strain. The first term on the right-hand side of Eq.~\eqref{eq:kocks} is a storage rate, while the second term is the depletion rate of stored dislocations. This equation does not conform with time reversal and reflection symmetries, as seen immediately if one reverses the strain rate which, strictly speaking, is a tensor. While one may circumvent this problem by replacing $\epsilon$ with $|\epsilon|$, this introduces a mathematical singularity at $\epsilon = 0$ which cannot be correct for physically well-posed and predictive evolution equations. These problems with the conventional literature suggest a vital need for physical input in multiscale descriptions of large deformations in metallic materials.

Related to the question of energy balance in a deforming polycrystalline solid is the accounting for the Taylor-Quinney coefficient, or the fraction of input work expended in heating up the material. Heat is primarily generated and confined within the ASB, causing local material softening that further reduces the resistance to dislocation glide, creating a positive feedback mechanism that results in more severe deformation. In spite of available experimental measurements~\citep[e.g.,][]{farren_1925,taylor_1934,hartley_1987,marchand_1988,duffy_1992,rittel_2017}, the accurate prediction of the thermal response of a plastically deforming material is of practical importance and remains an open question~\citep[e.g.,][]{zehnder_1991,rosakis_2000,benzerga_2005,longere_2008a,longere_2008b,stainier_2010,zaera_2013,anand_2015,luscher_2018}. Plasticity theories that directly address the issue of energy balance are most promising in this respect.

We recently proposed a minimal, thermodynamic description of DRX in polycrystalline solids during adiabatic shear localization \citep{lieou_2018}, based on the Langer-Bouchbinder-Lookman (LBL) theory of dislocation plasticity \citep{langer_2010,langer_2015}. The LBL theory directly addresses the issues with conventional dislocation theories, through the notion of a thermodynamically-defined effective temperature that describes the configurational state of the material in question, and the use of well-posed evolution equations for internal state variables. The fact that the LBL theory provides an accurate fit to strain hardening in copper over eight decades of strain rate with minimal assumptions \citep{langer_2015}, among other recent applications \citep[e.g.,][]{langer_2016,langer_2017a,langer_2017b,le_2018}, attests to its usefulness and predictive capabilities. In \citet{lieou_2018}, we augmented the LBL theory with a state variable for the grain boundary density, or the grain boundary area per unit volume. A very generic assumption for the interaction between grain boundaries and dislocations -- that the interaction is proportional to their respective densities -- immediately produces recrystallized grains in the ASB, and provides a good fit to experimental measurements in ultrafine-grained titanium, in a proof-of-principle calculation. DRX is seen to be an entropic effect; under severe loading conditions, the material forgoes dislocations in favor for an increased grain boundary density, the configuration which minimizes the free energy. Following the simple shear calculation in that manuscript, a natural extension is the implementation of the LBL-DRX theory in a simulation framework appropriate for more complex geometries and loading conditions in solid mechanics experiments and practical problems.

The present paper is devoted to a finite-element implementation and verification of the LBL-DRX theory in simulations; the rest of this paper is organized as follows. In Section \ref{sec:2} we present an overview of the LBL-DRX theory of dislocation plasticity and dynamic recrystallization, and discuss the physical basis of the thermodynamic approach. We also propose a simple way to compute the Taylor-Quinney coefficient of a deforming polycrystalline material. We describe in detail the computational framework in Section \ref{sec:3}, and the experiment on the 316L stainless steel in Section \ref{sec:4}. Section \ref{sec:5} summarizes the computational results and demonstrates good agreement with experiments. We conclude with a brief summary in Section \ref{sec:6}.

A list of symbols used throughout the paper is included in Table \ref{tab:sym} for the reader's convenience.

\begin{table}
\scriptsize
\begin{center}
\caption{\label{tab:sym}List of mathematical symbols}
\begin{tabular}{ll}
\hline
Symbol(s) & Meaning or definition \\
\hline\hline
$\sigma_{ij}$, $s_{ij}$ & Total and deviatoric stress tensors \\
$\dot{\epsilon}_{ij}$, $\dot{e}_{ij}$ & Total and deviatoric strain rate \\
$\dot{\epsilon}_{ij}^{\text{el}}$, $\dot{e}_{ij}^{\text{el}}$ & Total and deviatoric elastic strain rate \\
$\dot{\epsilon}_{ij}^{\text{pl}}$, $\dot{e}_{ij}^{\text{pl}}$ & Total and deviatoric plastic strain rate \\
$\bar{s}$, $\dot{\bar{e}}^{\text{pl}}$ & Stress and plastic strain rate invariants \\
$\bar{\rho}_M$ & Mass density\\
$\mu$ & Shear modulus\\
$\lambda$ & First Lam\'e parameter\\
$\nu$ & Poisson's ratio\\
$\mu_0$, $D_0$, $T_0$ & Parameters in shear modulus\\
$\beta$ & Taylor-Quinney factor\\
$c_v$, $c_p$ & Specific heat capacity, per unit volume and per unit mass\\
$c_0$, $c_1$ & Parameters in heat capacity $c_p$\\
$d$ & Grain size\\
$a$ & Atomic length scale \\
$v$ & Average dislocation speed\\
$\tau$ & Atomic vibration time scale\\
$e_P$ & Dislocation depinning energy barrier\\
$T$, $\theta$ & Thermal temperature, in Kelvins and energy units ($\theta = k_B T$)\\
$s_T$ & Taylor stress barrier\\
$\alpha_T$, $\mu_T$ & Taylor parameter, and effective shear modulus ($\mu_T = \alpha_T \mu$)\\
$q$ & Dimensionless plastic strain rate ($= 2 \tau \dot{\bar{e}}^{\text{pl}}$)\\
$\bar{\nu}$ & Dimensionless quantity $\bar{s} / s_T$\\
$e_D$ & Typical dislocation formation energy\\
$e_G$, $\tilde{e}_G$ & Typical grain boundary energy and its rescaled version ($\tilde{e}_G = e_G / e_D$)\\
$e_N$, $\tilde{e}_N$ & Typical dislocation-grain boundary interaction energy and its rescaled version ($\tilde{e}_N = e_N / e_D$)\\
$\rho$, $\tilde{\rho}$ & Dislocation density and its dimensionless version ($\tilde{\rho} = a^2 \rho$) \\
$\xi$, $\tilde{\xi}$ & Grain boundary density and its dimensionless version ($\tilde{\xi} = a \xi$)\\
$\chi$, $\tilde{\chi}$ & Effective temperature in energy units and its dimensionless version ($\tilde{\chi} = \chi / e_D$) \\
$\chi_0$, $\tilde{\chi}_0$ & Steady-state effective temperature and its dimensionless version ($\tilde{\chi}_0 = \chi_0 / e_D$) \\
$\tilde{\rho}^{\text{ss}}$, $\tilde{\xi}^{\text{ss}}$, $d^{\text{ss}}$ & Steady-state dimensionless dislocation density, dimensionless grain boundary density, and grain size \\
$U_C$, $S_C$ & Configurational energy and entropy\\
$U_K$, $S_K$ & Kinetic-vibrational energy and entropy\\
$U_{\text{tot}}$ & Total energy\\
$U_D$, $U_G$, $U_{\text{int}}$ & Energy density of dislocations, grain boundaries, and their interaction\\
$S_D$, $S_G$ & Entropy density of dislocations and grain boundaries\\
$K$ & Thermal transport coefficient between configurational and kinetic-vibrational degrees of freedom\\
$\kappa_1$, $\kappa_0$, $\kappa_r$ & Dislocation storage parameters\\
$\kappa_2$ & Disorder storage parameter\\
$\kappa_d$ & Grain boundary storage parameter\\
$q_r$ & Strain hardening parameter\\
\hline
\end{tabular}
\end{center}
\end{table}

\section{Thermodynamic theory of dislocation plasticity and dynamic recrystallization: an overview}
\label{sec:2}

In this section, we provide an overview of the LBL theory of dislocations and the recent extension we developed to describe dynamic recrystallization. The LBL theory \citep{langer_2010,langer_2015} provides a simple, minimal description of polycrystalline plasticity, consistent with the laws of thermodynamics. The recent extension of the theory to describe DRX, and a proof-of-principle calculation, are documented in our recent paper \citep{lieou_2018}.

\subsection{Kinematics and elasto-viscoplasticity}
\label{sec:2_1}

Let $\sigma_{ij}$ and $\dot{\epsilon}_{ij}$ denote the Cauchy stress and total strain rate tensors, and let $s_{ij}$ and $\dot{e}_{ij}$ denote their deviatoric counterparts. These are related to each other by
\begin{equation}
 s_{ij} = \sigma_{ij} - \dfrac{1}{3} \sigma_{kk} \delta_{ij}, \quad \dot{e}_{ij} = \dot{\epsilon}_{ij} - \dfrac{1}{3} \dot{\epsilon}_{kk} \delta_{ij} .
\end{equation}
(The repeated indices indicate the Einstein summation convention.) In polycrystalline metals, where the elastic strain is expected to be small, we decompose the total strain rate tensor additively into elastic and plastic parts, $\dot{\epsilon}_{ij}^{\text{el}}$ and $\dot{\epsilon}_{ij}^{\text{pl}}$:
\begin{equation}
 \dot{\epsilon}_{ij} = \dot{\epsilon}_{ij}^{\text{el}} + \dot{\epsilon}_{ij}^{\text{pl}} .
\end{equation}
Plastic incompressibility, or the notion that plastic deformation preserves volume, implies that the plastic strain-rate tensor is trace-free, or
\begin{equation}
 \dot{e}_{ij}^{\text{pl}} = \dot{\epsilon}_{ij}^{\text{pl}}.
\end{equation}

Assuming isotropic elasticity, the total stress rate is then given by
\begin{equation}\label{eq:sigma}
 \dot{\sigma}_{ij} = 2 \mu (\dot{\epsilon}_{ij} - \dot{\epsilon}_{ij}^{\text{pl}} ) + \lambda \dot{\epsilon}_{kk} \delta_{ij} .
\end{equation}
Here, $\mu$ is the shear modulus, and $\lambda$ is the first Lam\'e parameter, related to $\mu$ and the Poisson ratio $\nu$ by $\lambda = 2 \mu \nu / (1 - 2 \nu)$.

In this study, we assume that the shear modulus is temperature-dependent:
\begin{equation}\label{eq:mu}
 \mu = \mu_0 - \dfrac{D_0}{\exp (T / T_0) - 1} ,
\end{equation}
where $\mu_0$ and $D_0$ are parameters with the dimensions of stress, and $T_0$ is a reference temperature. We also assume that the Poisson ratio $\nu$ is a constant, and that $\lambda$ varies with the temperature accordingly.

\subsection{Dislocation motion}
\label{sec:2_2}

Define the deviatoric stress invariant
\begin{equation}
 \bar{s} = \sqrt{\dfrac{1}{2} s_{ij} s_{ij}} .
\end{equation}
Dislocation motion is governed by the Orowan relation, which says that the plastic strain rate is proportional to the dislocation density or dislocation line length per unit volume $\rho$, their average velocity $v$, and some atomic length scale $a$:
\begin{equation}\label{eq:orowan}
 \dot{\epsilon}_{ij}^{\text{pl}} = \dfrac{\rho}{2} \dfrac{s_{ij}}{\bar{s}} a v ,
\end{equation}
The Orowan relation, as written in Eq.~\eqref{eq:orowan}, assumes isotropic plasticity and co-directionality of plastic strain rate with the deviatoric stress. This is a fairly reasonable simplification in a polycrystalline material, where the crystal orientation varies between adjacent grains. In conventional literature, $a$ is usually the Burgers vector; we however take the view that $a$ is an atomic length scale, and absorb any uncertainties into the time scale associated with the velocity $v$.

A dislocation moves when it hops from one pinning site to another. The distance $l$ between pinning sites is related to the dislocation density $\rho$ by $l = 1 / \sqrt{\rho}$. We take the view that depinning is a thermally activated process with energy barrier $e_P$ and stress barrier $s_T$, so that the pinning time $\tau_P$ is given by
\begin{equation}\label{eq:tau_P}
 \dfrac{1}{\tau_P} = \dfrac{1}{\tau} \exp \left( - \dfrac{e_P}{\theta} e^{-s / s_T} \right) .
\end{equation}
Here, $\tau \sim 10^{-12}$ s, being the only relevant atomic time scale, is on the order of the inverse Debye frequency; we absorb any uncertainties in the length scale $b$ into here. $\theta = k_B T$ is the thermal temperature $T$ in energy units, with $k_B$ being the Boltzmann constant. The stress barrier $s_T$ equals the shear stress needed to unpin a dislocation and move it by a fraction of the length scale $a$ when the average separation between dislocations is $l = 1  /\sqrt{\rho}$. The shear strain associated with this operation is a fraction of the quantity $a / l = a \sqrt{\rho}$, so that it is given by the Taylor expression $s_T = \mu_T a \sqrt{\rho}$, where $\mu_T = \alpha_T \mu$ with $\alpha_T$ being on the order of $0.1$. Combining Eqs.~\eqref{eq:orowan} and \eqref{eq:tau_P}, the expression for the plastic strain rate is
\begin{equation}
 \dot{\epsilon}_{ij}^{\text{pl}} = \dfrac{\sqrt{\tilde{\rho}}}{2 \tau} \dfrac{s_{ij}}{\bar{s}} \exp \left( - \dfrac{e_P}{\theta} e^{-s / s_T} \right),
\end{equation}
which conveniently defines the dimensionless dislocation density $\tilde{\rho} = a^2 \rho$. It is useful to define the dimensionless plastic strain rate
\begin{equation}\label{eq:q}
 q \equiv 2 \tau \dot{\bar{\epsilon}}^{\text{pl}} = \sqrt{\tilde{\rho}} \exp \left[ - \dfrac{e_P}{\theta} e^{- \bar{s} / ( \mu_T \sqrt{\tilde{\rho}})} \right] .
\end{equation}
from the strain rate invariant $\dot{\bar{\epsilon}}^{\text{pl}} \equiv \sqrt{(1/2) \dot{\epsilon}_{ij}^{\text{pl}} \dot{\epsilon}_{ij}^{\text{pl}}}$.

\subsection{Nonequilibrium thermodynamics and steady-state defect densities}
\label{sec:2_3}

One of the most important aspects of the LBL theory of dislocation plasticity is the compliance with the laws of thermodynamics, based upon which the steady-state defect densities and the evolution of state variables are derived. This is the place where the present theory diverges from traditional descriptions of polycrystalline plasticity and dynamic recrystallization. The deforming polycrystalline material is by definition in a nonequilibrium state because of the nonzero external work rate arising from deformation itself. The configurational degrees of freedom -- those pertaining to the positions of atoms -- fall out of equilibrium with the kinetic-vibrational degrees of freedom pertaining to the atoms' thermal motion, whose time scale given by the Debye frequency is often many orders of magnitude above the strain rate. As such, we partition the total energy density $U_{\text{tot}}$ and entropy density $S_{\text{tot}}$ into configurational (C) and kinetic-vibrational (K) contributions:
\begin{equation}
 U_{\text{tot}} = U_C + U_K ; \quad S_{\text{tot}} = S_C + S_K.
\end{equation}
Dislocations and grain boundaries (GBs) clearly belong to the configurational degrees of freedom. Denote by $\xi$ the GB density, or the GB area per unit volume, and its dimensionless counterpart $\tilde{\xi} = a \xi$. (The characteristic grain size $d$ is related to the GB density by $d = 1 / \xi$.) The configurational energy and entropy densities $U_C$ and $U_K$ can then be written as
\begin{eqnarray}
 U_C (S_C, \tilde{\rho}, \tilde{\xi}) &=& U_D (\tilde{\rho}) + U_G (\tilde{\xi}) + U_{\text{int}} (\tilde{\rho}, \tilde{\xi}) + U_1 (S_1) ; \\
 S_C (U_C, \tilde{\rho}, \tilde{\xi}) &=& S_D (\tilde{\rho}) + S_G (\tilde{\xi}) + S_1 (U_1) .
\end{eqnarray}
$U_D$ and $S_D$ are the energy and entropy densities associated with dislocations; their counterparts for GBs are $U_G$ and $S_G$. $U_1$ and $S_1$ are the energy and entropy densities of all other configurational degrees of freedom. We implicitly assume that the contributions of dislocations and GBs to the entropy are independent, while there is a contribution $U_{\text{int}}$ from the interaction between dislocations and GBs to the total energy density. Define the \textit{effective} temperature
\begin{equation}
 \chi \equiv \dfrac{\partial U_C}{\partial S_C} .
\end{equation}
The first law of thermodynamics says that
\begin{eqnarray}\label{eq:U_tot}
 \dot{U}_{\text{tot}} &=& \sigma_{ij} \dot{\epsilon}_{ij} = \dot{U}_C + \dot{U}_K \\ &=& \chi \dot{S}_C + \left( \dfrac{\partial U_C}{\partial t} \right)_{S_C,\tilde{\rho},\tilde{\xi}} + \left( \dfrac{\partial U_C}{\partial \tilde{\rho}} \right)_{S_C,\tilde{\xi}} \dot{\tilde{\rho}} + \left( \dfrac{\partial U_C}{\partial \tilde{\xi}} \right)_{S_C,\tilde{\rho}} \dot{\tilde{\xi}} + \theta \dot{S}_K . ~~~~~
\end{eqnarray}
Because deformation at constant defect densities $\tilde{\rho}$, $\tilde{\xi}$ and configurational entropy $S_C$ is by definition elastic \citep{bouchbinder_2009b}, the elastic work $(\partial U_C / \partial t)_{S_C, \tilde{\rho}, \tilde{\xi}} = \sigma_{ij} \dot{\epsilon}_{ij}^{\text{el}}$ cancels out of both sides of Eq.~\eqref{eq:U_tot}, so that
\begin{equation}\label{eq:first_law}
 \sigma_{ij} \dot{\epsilon}_{ij}^{\text{pl}} = \chi \dot{S}_C + \left( \dfrac{\partial U_C}{\partial \tilde{\rho}} \right)_{S_C,\tilde{\xi}} \dot{\tilde{\rho}} + \left( \dfrac{\partial U_C}{\partial \tilde{\xi}} \right)_{S_C,\tilde{\rho}} \dot{\tilde{\xi}} + \theta \dot{S}_K .
\end{equation}

Move on to the second law of thermodynamics, according to which
\begin{equation}
 \dot{S}_{\text{tot}} = \dot{S}_C + \dot{S}_K \geq 0.
\end{equation}
Eliminating $\dot{S}_C$ using the energy balance equation, one finds
\begin{equation}\label{eq:second_law}
 \sigma_{ij} \dot{\epsilon}_{ij}^{\text{pl}} - \left( \dfrac{\partial U_C}{\partial \tilde{\rho}} \right)_{S_C} \dot{\tilde{\rho}} - \left( \dfrac{\partial U_C}{\partial \tilde{\xi}} \right)_{S_C} \dot{\tilde{\xi}} + (\chi - \theta) \dot{S}_K \geq 0.
\end{equation}
The Coleman-Noll argument \citep{coleman_1963} stipulates non-negativity of each independently variable term in this inequality. This is automatically satisfied for the plastic work rate $\sigma_{ij} \dot{\epsilon}_{ij}^{\text{pl}} = s_{ij} \dot{e}_{ij}^{\text{pl}}$ according to Eq.~\eqref{eq:orowan}. The constraint $(\chi - \theta) \dot{S}_K \geq 0$ will be discussed in Section \ref{sec:2_5} in connection with the Taylor-Quinney coefficient which determines the fraction of plastic work dissipated as heat. For now, we are left with
\begin{equation}\label{eq:U_C_constraint}
 - \left( \dfrac{\partial U_C}{\partial \tilde{\rho}} \right)_{S_C} \dot{\tilde{\rho}} \geq 0; \quad - \left( \dfrac{\partial U_C}{\partial \tilde{\xi}} \right)_{S_C} \dot{\tilde{\xi}} \geq 0 .
\end{equation}
These inequalities say that the time rates of change of the defect densities $\tilde{\rho}$ and $\tilde{\xi}$ change sign when $U_C$ at constant $S_C$ is a minimum. Once we write explicitly that
\begin{eqnarray}
 \left( \dfrac{\partial U_C}{\partial \tilde{\rho}} \right)_{S_C} &=& \dfrac{\partial U_D}{\partial \tilde{\rho}} + \dfrac{\partial U_{\text{int}}}{\partial \tilde{\rho}} - \chi \dfrac{\partial S_D}{\partial \tilde{\rho}} \equiv \dfrac{\partial F_C}{\partial \tilde{\rho}} ; \\
 \left( \dfrac{\partial U_C}{\partial \tilde{\xi}} \right)_{S_C} &=& \dfrac{\partial U_G}{\partial \tilde{\xi}} + \dfrac{\partial U_{\text{int}}}{\partial \tilde{\xi}}- \chi \dfrac{\partial S_G}{\partial \tilde{\xi}} \equiv \dfrac{\partial F_C}{\partial \tilde{\xi}} ,
\end{eqnarray}
where 
\begin{equation}\label{eq:F_C}
 F_C (\tilde{\rho}, \tilde{\xi}) = U_D (\tilde{\rho}) + U_G (\tilde{\xi}) + U_{\text{int}} (\tilde{\rho}, \tilde{\xi}) - \chi (S_D (\tilde{\rho}) + S_G (\tilde{\xi}) )
\end{equation}
is the configurational free energy density, the requirement in \eqref{eq:U_C_constraint} is immediately seen to amount to the dynamic minimization of the configurational free energy.

A very generic assumption is that the energy density of dislocations $U_D$ and GBs $U_G$ increase linearly with the dislocation and GB densities, and that the interaction energy $U_{\text{int}}$ is bilinear in the defect densities; i.e.,
\begin{equation}
 U_D = \dfrac{e_D \tilde{\rho}}{a^3}, \quad U_G = \dfrac{e_G \tilde{\xi}}{a^3}, \quad U_{\text{int}} = \dfrac{e_N \, \tilde{\rho} \, \tilde{\xi}}{a^3} .
\end{equation}
This defines the characteristic formation energies $e_D$ and $e_G$ for a dislocation line of length $a$ and a GB of area $a^2$, respectively, and the energy scale $e_N$. The length scale $a$ is therefore the minimum average separation between dislocation lines and between GBs, or the minimum length for which dislocation and GB densities are meaningful quantities, and should be roughly 10-20 atomic spacings. The entropies $S_D$ and $S_G$ can be computed by a simple counting argument detailed in, for example, \citet{lieou_2018}, with the result
\begin{equation}
 S_D (\tilde{\rho}) = \dfrac{1}{a^3} ( - \tilde{\rho} \ln \tilde{\rho} + \tilde{\rho} ), \quad S_G (\tilde{\xi}) = \dfrac{1}{a^3} ( - \tilde{\xi} \ln \tilde{\xi} + \tilde{\xi} ).
\end{equation}
As such, the steady-state defect densities are given by
\begin{eqnarray}
 \label{eq:rho_ss} \tilde{\rho}^{\text{ss}} &=& \exp \left( - \dfrac{e_D + e_N \tilde{\xi}}{\chi} \right) ; \\
 \label{eq:xi_ss} \tilde{\xi}^{\text{ss}} &=& \exp \left( - \dfrac{e_G + e_N \tilde{\rho}}{\chi} \right) .
\end{eqnarray}
It is seen that whenever $e_D > e_G$ and $e_N > 0$, dynamically recrystallized grains with depleted dislocations correspond to the steady state. Of course, there needs to be a pathway, i.e., large enough strain rate, for the polycrystalline material to reach this DRX state to begin with.

It is convenient to rescale the effective temperature $\chi$ and the defect energies $e_G$ and $e_N$ by the dislocation formation energy $e_D$:
\begin{equation}
 \tilde{\chi} \equiv \chi / e_D, \quad \tilde{e}_G \equiv e_G / e_D, \quad \tilde{e}_N \equiv e_N / e_D .
\end{equation}
Then
\begin{eqnarray}
 \label{eq:rho_ss2} \tilde{\rho}^{\text{ss}} &=& \exp \left( - \dfrac{1 + \tilde{e}_N \tilde{\xi}}{\tilde{\chi}} \right) ; \\
 \label{eq:xi_ss2} \tilde{\xi}^{\text{ss}} &=& \exp \left( - \dfrac{\tilde{e}_G + \tilde{e}_N \tilde{\rho}}{\tilde{\chi}} \right) .
\end{eqnarray}

\subsection{Evolution of defect densities and the effective temperature}
\label{sec:2_4}

Energy is stored in dislocations and grain boundaries that are formed over the course of deformation. In order to express a direct connection between the rate at which mechanical work is done on the material and the rates at which defects are created or annihilated, the rates of change of the defect densities, $\dot{\tilde{\rho}}$ and $\dot{\tilde{\xi}}$, are manifestly proportional to the plastic work rate $\sigma_{ij} \dot{\epsilon}_{ij}^{\text{pl}} = s_{ij} \dot{e}_{ij}^{\text{pl}}$, which is the only relevant scalar invariant with the dimensions of energy per unit volume per time. The equation for the dislocation density evolution towards the steady-state value $\tilde{\rho}^{\text{ss}}$ reads
\begin{equation}\label{eq:rho}
 \dot{\tilde{\rho}} = \kappa_1 \dfrac{\sigma_{ij} \dot{\epsilon}_{ij}^{\text{pl}}}{\bar{\nu}^2 \mu_T} \left( 1 - \dfrac{\tilde{\rho}}{\tilde{\rho}^{\text{ss}}} \right) .
\end{equation}
Here, the quantity 
\begin{equation}\label{eq:barnu}
 \bar{\nu} \equiv \dfrac{\bar{s}}{\mu_T \sqrt{\tilde{\rho}}} = \ln \left( \dfrac{e_P}{\theta} \right) - \ln \left[ \ln \left( \dfrac{\sqrt{\tilde{\rho}}}{q} \right) \right],
\end{equation}
where $q$ is the dimensionless strain rate defined in Eq.~\eqref{eq:q}, controls the strain-hardening rate. The $\bar{\nu}^{-2}$ dependence on the right-hand side of Eq.~\eqref{eq:rho} can be derived by computing the hardening rate at the onset of plasticity, when $\bar{s} \approx s_T = \mu_T \sqrt{\tilde{\rho}}$, $\dot{\epsilon}_{ij} \approx \dot{\epsilon}_{ij}^{\text{pl}}$, and $\tilde{\rho} \ll \tilde{\rho}^{\text{ss}}$~\citep{langer_2010,langer_2015}. $\kappa_1$ is a storage factor which increases with the strain rate, and increases with decreasing grain size because grain corners serve as a source of dislocations; as in \citep{lieou_2018}, we assume the form
\begin{equation}
 \kappa_1 (d, q) = \kappa_0 + \dfrac{\kappa_r}{\sqrt{d}} \left( 1 + \dfrac{q}{q_r} \right) ,
\end{equation}
where $q_r$ determines the onset of rate hardening, and $\kappa_0$, $\kappa_r$ are hardening parameters.

The evolution equation for the GB density is 
\begin{equation}\label{eq:xi}
 \dot{\tilde{\xi}} = \kappa_d \dfrac{\sigma_{ij} \dot{\epsilon}_{ij}^{\text{pl}}}{\mu_T} \tilde{\xi} \left( 1 - \dfrac{\tilde{\xi}}{\tilde{\xi}^{\text{ss}}} \right) ,
\end{equation}
where $\kappa_d$ is a dimensionless GB storage parameter. Note that in contrast to our earlier work \citep{lieou_2018}, here we inserted an overall factor of $\tilde{\xi}$ on the right hand side. This factor appears to be necessary to ensure timely recrystallization going from micron-sized grains to grains with diameter $d \sim 100$ nm. If one prefers to track the evolution of the grain size $d = a / \xi$ as opposed to $\tilde{\xi}$, the evolution equation is
\begin{equation}\label{eq:d}
\dot{d} = \kappa_d \dfrac{\sigma_{ij} \dot{\epsilon}_{ij}^{\text{pl}}}{\mu_T} \left( d^{\text{ss}} - d \right),
\end{equation}
where $d^{\text{ss}} = a / \tilde{\xi}^{\text{ss}}$ is the steady-state grain size.

The effective temperature $\tilde{\chi}$ increases as deformation induces configurational disorder, and saturates at some $\tilde{\chi}_0$:
\begin{equation}\label{eq:chi}
 \dot{\tilde{\chi}} = \dfrac{\kappa_2}{\mu_T} \sigma_{ij} \dot{\epsilon}_{ij}^{\text{pl}} \left( 1 - \dfrac{\tilde{\chi}}{\tilde{\chi}_0} \right) ,
\end{equation}
with $\kappa_2$ being a dimensionless parameter. At saturation, $\tilde{\rho}^{\text{ss}} \approx e^{-1 / \tilde{\chi}_0}$, and the average separation between dislocation lines should be about $10 a$, in the spirit of the Lindemann melting criterion~\citep{lindemann_1910}; this gives $\tilde{\chi}_0 \approx 0.25$.

Finally, the true, thermal temperature increases at a rate proportional to the plastic work rate:
\begin{equation}\label{eq:theta}
 \dot{\theta} = k_B \dot{T} = \dfrac{\beta k_B}{c_v} \sigma_{ij} \dot{\epsilon}_{ij}^{\text{pl}} .
\end{equation}
In the adiabatic approximation, we neglect the flow of heat within the material, and between the material and the surroundings. This is valid as long as heat is generated more quickly than the speed of heat conduction, and is a good approximation at sufficiently high strain rates. The Taylor-Quinney coefficient $\beta$ is the fraction of plastic work converted into heat. $c_v$ is the heat capacity per unit volume of the material; it is related to the heat capacity per unit mass $c_p$ and the mass density $\bar{\rho}_M$ by $c_v = c_p \bar{\rho}_M$; in this study we assume temperature dependence of the form
\begin{equation}
 c_p (T) = c_0 + c_1 T ,
\end{equation}
where $c_0$ and $c_1$ are constants.

\subsection{Taylor-Quinney coefficient}
\label{sec:2_5}

Incidentally, the thermodynamic description outlined in Section \ref{sec:2_3} provides a simple estimate of the Taylor-Quinney coefficient $\beta$, a long-standing challenge in the materials science and solid mechanics communities \citep[e.g.,][]{zehnder_1991,rosakis_2000,benzerga_2005,longere_2008a,longere_2008b,stainier_2010,zaera_2013,anand_2015,luscher_2018} despite a vast amount of experimental efforts~\citep[e.g.,][]{farren_1925,taylor_1934,hartley_1987,marchand_1988,duffy_1992,rittel_2017}. The thermodynamic constraint $(\chi - \theta) \dot{S}_K \geq 0$, a direct consequence of the second-law inequality \eqref{eq:second_law}, stipulates that
\begin{equation}\label{eq:theta2}
 c_v \dot{\theta} = \theta \dot{S}_K = - K \left( 1 - \dfrac{\chi}{\theta} \right) ,
\end{equation}
where $K$ is a non-negative thermal transport coefficient \citep{bouchbinder_2009b}. To calculate $K$, consider the nonequilibrium steady state, at which the effective temperature $\chi$, as well as the dislocation and GB densities $\tilde{\rho}$ and $\tilde{\xi}$, have reached their respective steady-state values, i.e., $\chi_0 \equiv e_D \tilde{\chi}_0$, and $\dot{\chi} = 0$, $\dot{\tilde{\rho}} = \dot{\tilde{\xi}} = 0$. Substitution of these and Eq.~\eqref{eq:theta2} into the first-law statement, Eq.~\eqref{eq:first_law}, gives
\begin{equation}
 K = \sigma_{ij} \dot{\epsilon}_{ij}^{\text{pl}} \dfrac{\theta}{\chi_0 - \theta} .
\end{equation}
Suppose that this also holds true beyond the nonequilibrium steady state, by virtue of consistency. Then
\begin{equation}
 c_v \dot{\theta} = \sigma_{ij} \dot{\epsilon}_{ij}^{\text{pl}} \dfrac{\chi - \theta}{\chi_0 - \theta} ,
\end{equation}
from which we directly read off the Taylor-Quinney coefficient
\begin{equation}
 \beta = \dfrac{\chi - \theta}{\chi_0 - \theta} .
\end{equation}
Because thermal fluctuations of energy $\theta = k_B T$ are insufficient to create dislocations and grain boundaries, $\theta \ll e_G \simeq e_D \lesssim \chi \lesssim \chi_0$. As such,
\begin{equation}
 \beta \approx \dfrac{\chi}{\chi_0} = \dfrac{\tilde{\chi}}{\tilde{\chi}_0} .
\end{equation}
The present argument says that the Taylor-Quinney coefficient, or the fraction of plastic work expended in heating up the material, is entirely controlled by the state of its configurational disorder, increasing towards unity as the deforming material approaches the nonequilibrium steady state, at which all of the input work is dissipated as heat.

\section{Computational method}
\label{sec:3}

To solve evolution equations, which include Eq.~\eqref{eq:sigma} for the stress, and Eqs.~\eqref{eq:rho}, \eqref{eq:d}, \eqref{eq:xi}, and \eqref{eq:theta} for the dislocation density $\tilde{\rho}$, grain size $d$, effective temperature $\tilde{\chi}$, and thermal temperature $T$, we implement the following implicit algorithm within the explicit branch of the finite element code ABAQUS \citep{abaqus_2014}. The symmetry and geometry of the hat-shaped sample permits the use of axisymmetric formulation where we represent the material by a vertical cross section, partitioned into quadrilateral elements with four nodes each, two degrees of freedom (radial and axial displacement) per node, and four independent components for the stress and strain rate tensors associated with each element (the $rr$, $\theta \theta$, $zz$, and $rz$ components).

The implicit algorithm is as follows. Denote by $\Lambda_{\alpha}$ the collection of state variables $\tilde{\rho}$, $d$, $\tilde{\chi}$, and $T$ at each element. At each time step $t$, we first perform an explicit update to make a best guess for the state variables at the next time step at $t + \Delta t$:
\begin{equation}\label{eq:num_trial}
 \Lambda_{\alpha}^* (t + \Delta t) = \Lambda_{\alpha}(t) + \Delta t \cdot \dot{\Lambda}_{\alpha} (\sigma_{ij} (t), \Lambda_{\beta} (t)) .
\end{equation}
If $t = 0$, we perform an explicit update to compute the stress value $\sigma_{ij}$ at the next time step. Otherwise, we perform the following Newton-type iterative algorithm to compute the stress. If $\Delta \epsilon_{ij}$ is the strain increment accrued through time $\Delta t$, define
\begin{equation}
 R_{ij} \left( \sigma_{kl} (t + \Delta t) \right) \equiv \sigma_{ij} (t + \Delta t) - \sigma_{ij} (t) - 2 \mu \left[ \Delta \epsilon_{ij} - \Delta t \, \dot{\epsilon}_{ij}^{\text{pl}} (t + \Delta t) \right] - \lambda \delta_{ij} \Delta \epsilon_{kk} .
\end{equation}
Note that $\dot{\epsilon}_{ij}^{\text{pl}} (t + \Delta t)$ is a function of $\sigma_{kl} (t + \Delta t)$ and the state variables at time $t + \Delta t$. $\sigma_{ij} (t + \Delta t)$ is then found by setting $R_{ij} \left( \sigma_{kl} (t + \Delta t) \right) = 0$. The iterative solution going from the $n$th to the $(n+1)$-st iteration is
\begin{equation}\label{eq:num_stress}
 \sigma_{ij}^{(n+1)} (t + \Delta t) = \sigma_{ij}^{(n)} (t + \Delta t) - (J^F_{kl,ij})^{-1} R_{kl} \left( \sigma_{kl}^{(n)} (t + \Delta t) \right),
\end{equation}
where
\begin{equation}
 J^F_{ij,kl} \equiv \dfrac{\partial R_{ij} \left( \sigma_{kl}^{(n)} (t + \Delta t) \right)}{\partial \sigma_{kl}^{(n)} (t + \Delta t)} = \delta_{ij,kl} + 2 \mu \Delta t \dfrac{\partial \dot{\epsilon}_{ij}^{\text{pl}} \left( \sigma_{kl}^{(n)} (t + \Delta t) \right)}{\partial \sigma_{kl}^{(n)} (t + \Delta t)}
\end{equation}
is the 4-by-4 Jacobian of the function $R_{ij} \left( \sigma_{kl}^{(n)} (t + \Delta t) \right)$ defined above (no summation over the four possible pairs of $kl$-indices here), and $\delta_{ij,kl} = 1$ when $(ij) = (kl)$ and 0 otherwise. Upon reaching sufficient accuracy for $\sigma_{ij} (t + \Delta t)$, we stop the iteration, and perform one final update for the state variables at time $t + \Delta t$:
\begin{equation}
 \Lambda_{\alpha} (t + \Delta t) = \Lambda_{\alpha} (t) + \Delta t \cdot \dot{\Lambda}_{\alpha} ( \sigma_{ij} (t + \Delta t), \Lambda_{\beta}^* (t + \Delta t) ),
\end{equation}
using the trial values $\Lambda_{\beta}^* (t + \Delta t)$ obtained from Eq.~\eqref{eq:num_trial} above, and the stress $\sigma_{ij} (t + \Delta t)$ from Eq.~\eqref{eq:num_stress}.

\section{Experiments}
\label{sec:4}

\begin{figure}
\begin{center}
\includegraphics[width=0.8\textwidth]{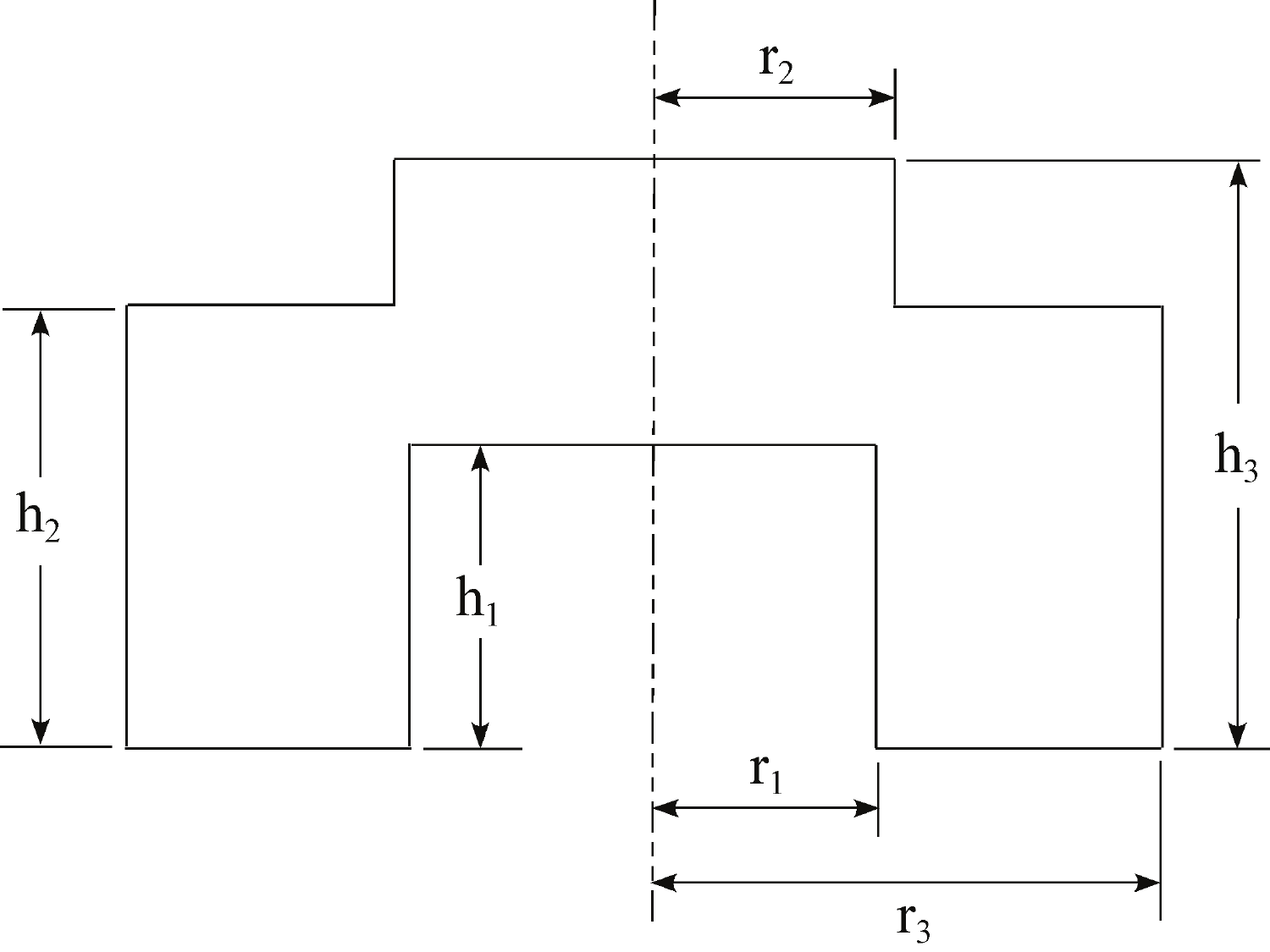}
\caption{\label{fig:tophat}Schematic drawing of the cross-section of the axisymmetric sample.}
\end{center}
\end{figure}

The cylindrical hat-shaped sample geometry, first developed by \citet{meyer_1985} and depicted here in Fig.~\ref{fig:tophat}, has been exploited in previous work \citep[e.g.,][]{bronkhorst_2006} to study the shear-dominated response of metallic materials, because of the oblique orientation of the shear band relative to the loading direction. Sample dimensions of the AISI 316L stainless steel specimen used in this study are tabulated in Table \ref{tab:hat}.

\begin{table}
\scriptsize
\begin{center}
\caption{\label{tab:hat}Dimensions of the hat-shaped 316L steel sample used in the present experiment and sketched in Fig.~\ref{fig:tophat}}
\begin{tabular}{ll}
\hline
Dimension variable corresponding to Fig.~\ref{fig:tophat} & Value (mm) \\
\hline\hline
$r_1$ & 2.095 \\
$r_2$ & 2.285 \\
$r_3$ & 4.320 \\
$h_1$ & 2.540 \\
$h_2$ & 3.430 \\
$h_3$ & 5.080 \\
\hline
\end{tabular}
\end{center}
\end{table}

\begin{figure}
\begin{center}
\includegraphics[width=0.6\textwidth]{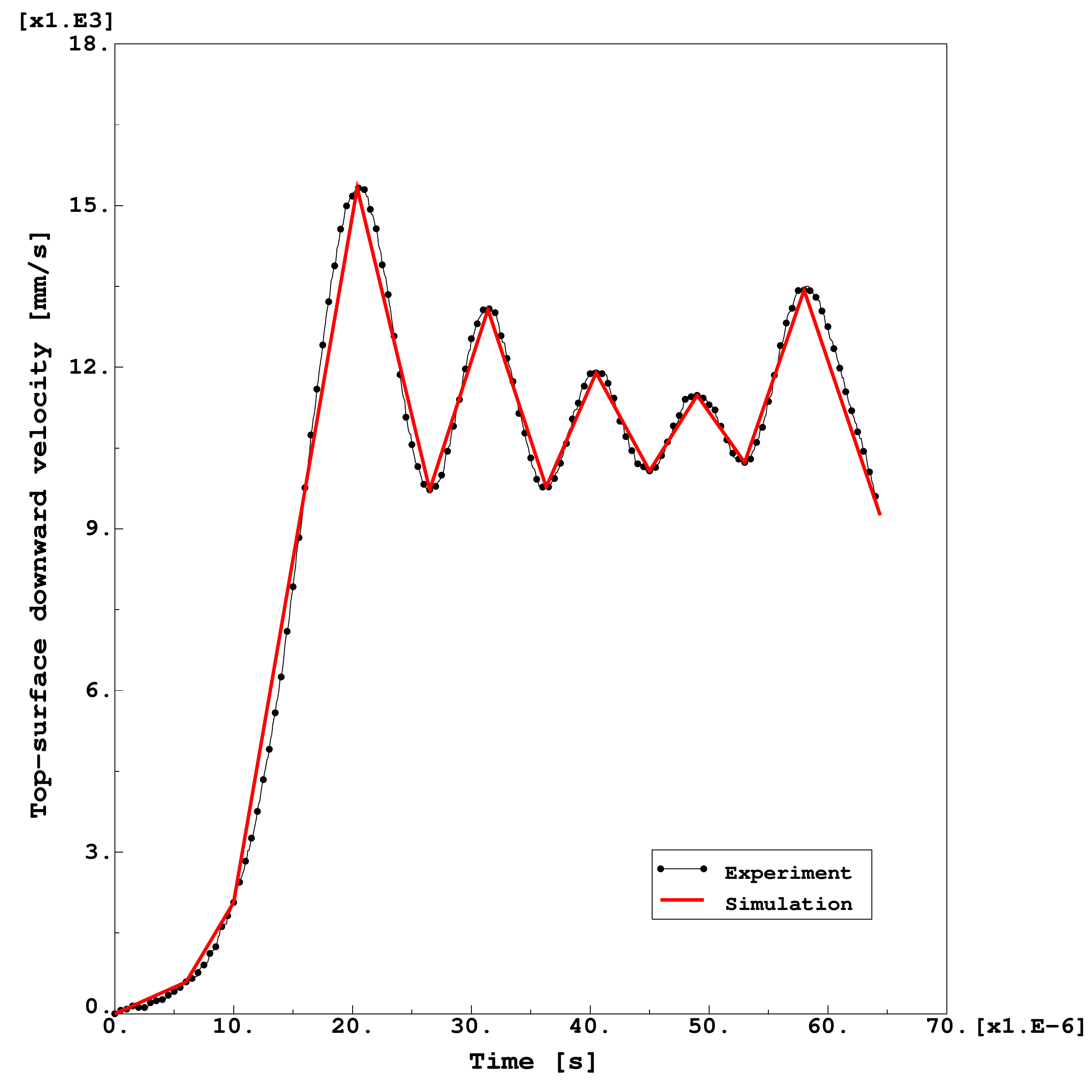}
\caption{\label{fig:B_velocity}Imposed top-surface downward velocity on the hat-shaped steel sample. The data points are recorded in the experiment, while the solid red line is the approximation used in our simulation as the top-surface boundary condition.}
\end{center}
\end{figure}

In the experiment that we consider in this manuscript and also described in \citet{mourad_2017}, a series of identical samples were loaded dynamically from the top, using a split-Hopkinson pressure bar test system. Steel collars were placed around the sample to avoid overdrive and to arrest the sample at pre-determined displacements. The tests were conducted at an initial temperature of $T = 298$ K and a breech gas pressure of 42 kPa. The striker bar length was 15.24 cm. The initial grain size of each sample was $d = 30 \mu$m. Fig.~\ref{fig:B_velocity} shows the downward velocity profile imposed at the top of each specimen.

\section{Model results}
\label{sec:5}

\begin{figure}
\begin{center}
\begin{subfigure}[b]{.45\textwidth}
\includegraphics[width=\textwidth]{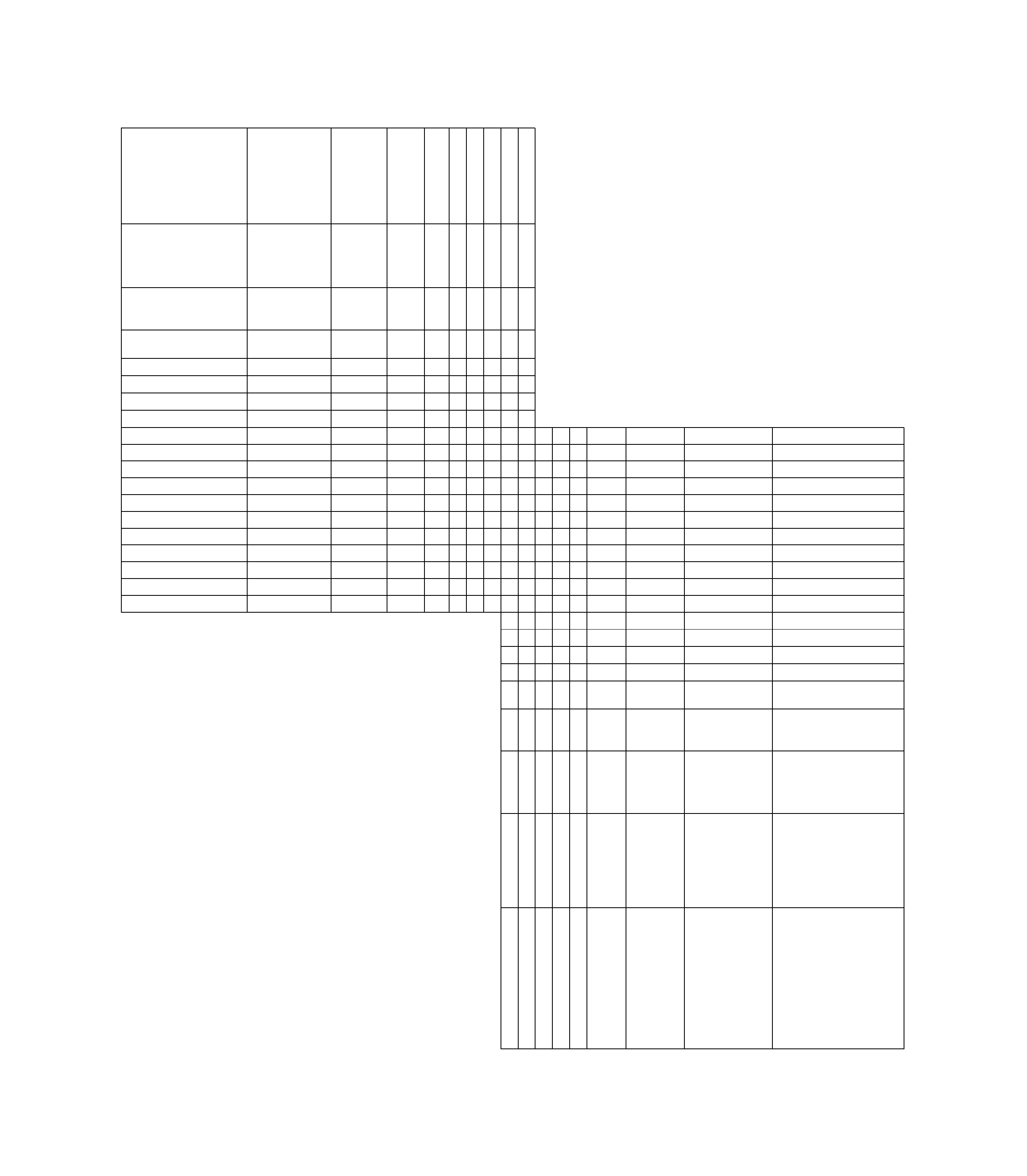}\caption{$h = 90 \mu$m}
\end{subfigure}
\begin{subfigure}[b]{.45\textwidth}
\includegraphics[width=\textwidth]{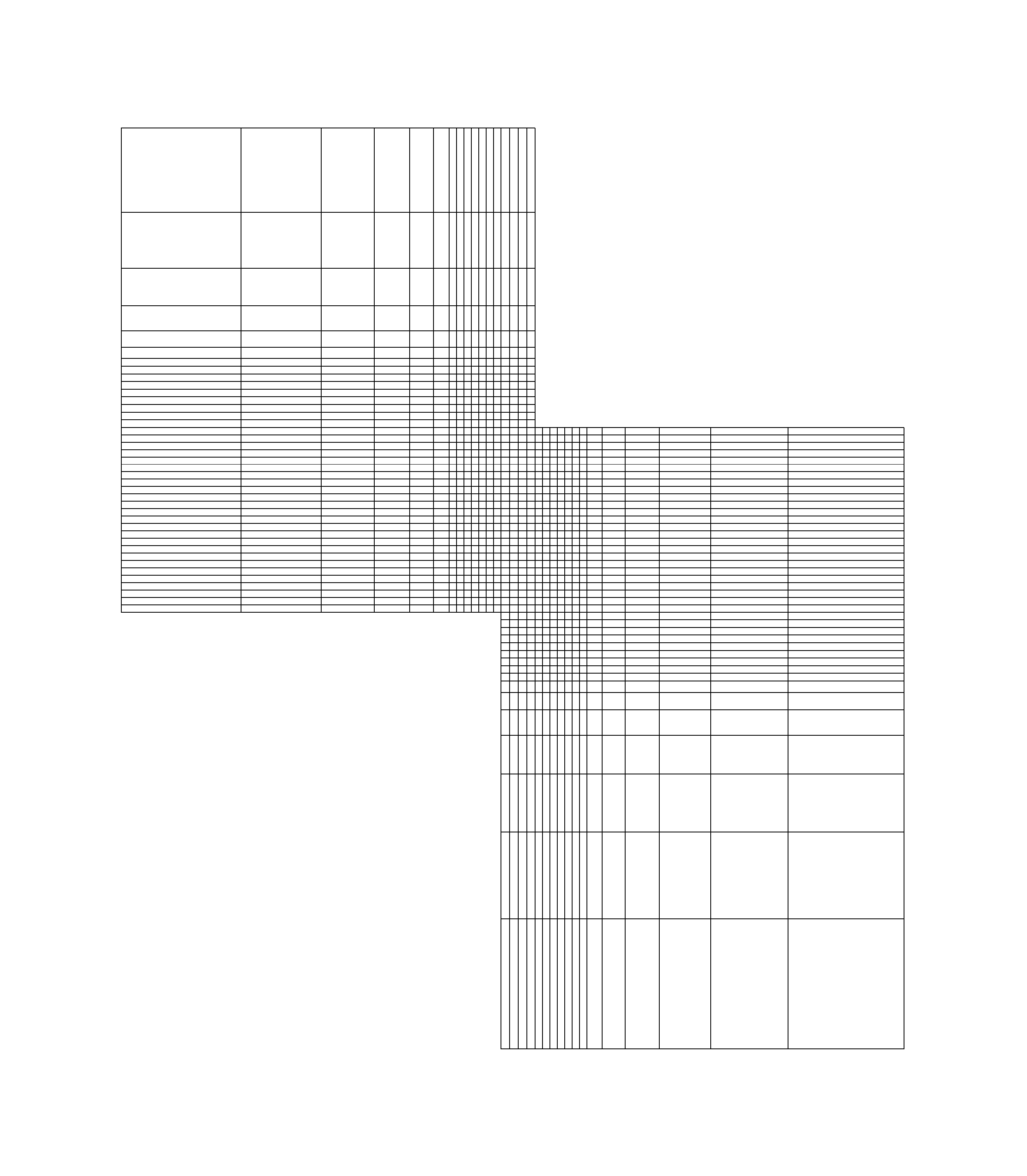}\caption{$h = 40 \mu$m}
\end{subfigure}
\caption{\label{fig:mesh}Two of the meshes used in the present study, with element size in the shear section being (a) $h = 90$ $\mu$m and (b) $ h = 40$ $\mu$m.}
\end{center}
\end{figure}

We present in this section the simulations results based on the traditional finite element method and the implicit algorithm presented in Section \ref{sec:3}. Three finite element meshes have been used in this study, with element sizes $h = 90$, 40, and 20 $\mu$m in the shear section; two of these are shown in Fig.~\ref{fig:mesh}. Note that the mesh itself introduces the length scale $h$ into the problem; without using more sophisticated sub-grid methods that constrain the shear-band width, $h$ limits the shear band width, and some length-related parameters, such as the atomic length scale $a$ over which one can define dislocation and GB densities, are presently $h$-dependent. The material parameters used in the present work are listed in Table \ref{tab:parameters}. For steel, many parameters are known. For example, its heat capacity, mass density, and the parameters associated with the shear modulus $\mu$ are well documented in the literature \citep[e.g.,][]{bronkhorst_2006,mourad_2017}; LBL theory parameters appropriate for steel, such as the depinning energy $e_P$, the ratio $\alpha_T = \mu_T / \mu$, and the storage coefficients $\kappa_1$ and $\kappa_2$, are documented in \citet{le_2018}. We only needed to adjust the parameters $\tilde{e}_N$ and $\tilde{e}_G$ for the ratios of the interaction energy and GB energy scales to the dislocation formation energy, and the GB energy storage parameter $\kappa_d$. We also account for the grain-size and strain-rate dependence of the storage parameter $\kappa_1$ through small adjustments in $q_r$ and $\kappa_r$ to keep $\kappa_1$ in the same ballpark as reported in the literature. In addition, we had to adjust the initial conditions for the dislocation density and effective temperature to provide a good fit to the stress-displacement curve, the only piece of experimental measurement directly available to us; we used $\tilde{\rho} (t = 0) = 1.7 \times 10^{-3}$ for the relatively large strain hardening at the initial stage, and $\tilde{\chi} (t = 0) = 0.16$.

\begin{table}
\scriptsize
\begin{center}
\caption{\label{tab:parameters}List of parameters and initial conditions}
\begin{tabular}{lll}
\hline
Parameter & Definition or meaning & Value \\
\hline\hline
$\bar{\rho}_M$ & Mass density & 7860 kg m$^{-3}$ \\
$\mu_0$ & Shear modulus parameter & 71.46 GPa \\
$D_0$ & Shear modulus parameter & 2.09 GPa \\
$T_0$ & Shear modulus parameter & 204 K \\
$\nu$ & Poisson's ratio & 0.3 \\
$c_0$ & Heat capacity parameter & 391.63 J kg$^{-1}$ K$^{-1}$ \\
$c_1$ & Heat capacity parameter & 0.237 J kg$^{-1}$ K$^{-2}$ \\
$a$ & Atomic length scale & 12, 1.8, 1 nm for $h = 90, 40, 20$ $\mu$m \\
$\tau$ & Atomic time scale & 1 ps \\
$e_P$ & Depinning energy barrier & $7.121 \times 10^{-18}$ J \\
$\alpha_T$ & Ratio $\mu / \mu_T$ & 0.0178 \\
$\tilde{e}_G$ & GB energy in units of dislocation energy $e_D$ & 0.2 \\
$\tilde{e}_N$ & GB-dislocation interaction in units of dislocation energy $e_D$ & 100 \\
$\tilde{\chi}_0$ & Steady-state effective temperature in units of $e_D$ & 0.25 \\
$\kappa_r$ & Dislocation storage rate parameter & $10^{-3}$ \\
$q_r$ & Rate hardening parameter & $2 \times 10^{-9}$ \\
$\kappa_0$ & Dislocation storage rate parameter & 7.5 \\
$\kappa_2$ & Effective temperature increase rate & 14.3 \\
$\kappa_d$ & Recrystallization rate parameter & 5 \\
\hline
\end{tabular}
\end{center}
\end{table}

\begin{figure}
\begin{center}
\includegraphics[width=0.6\textwidth]{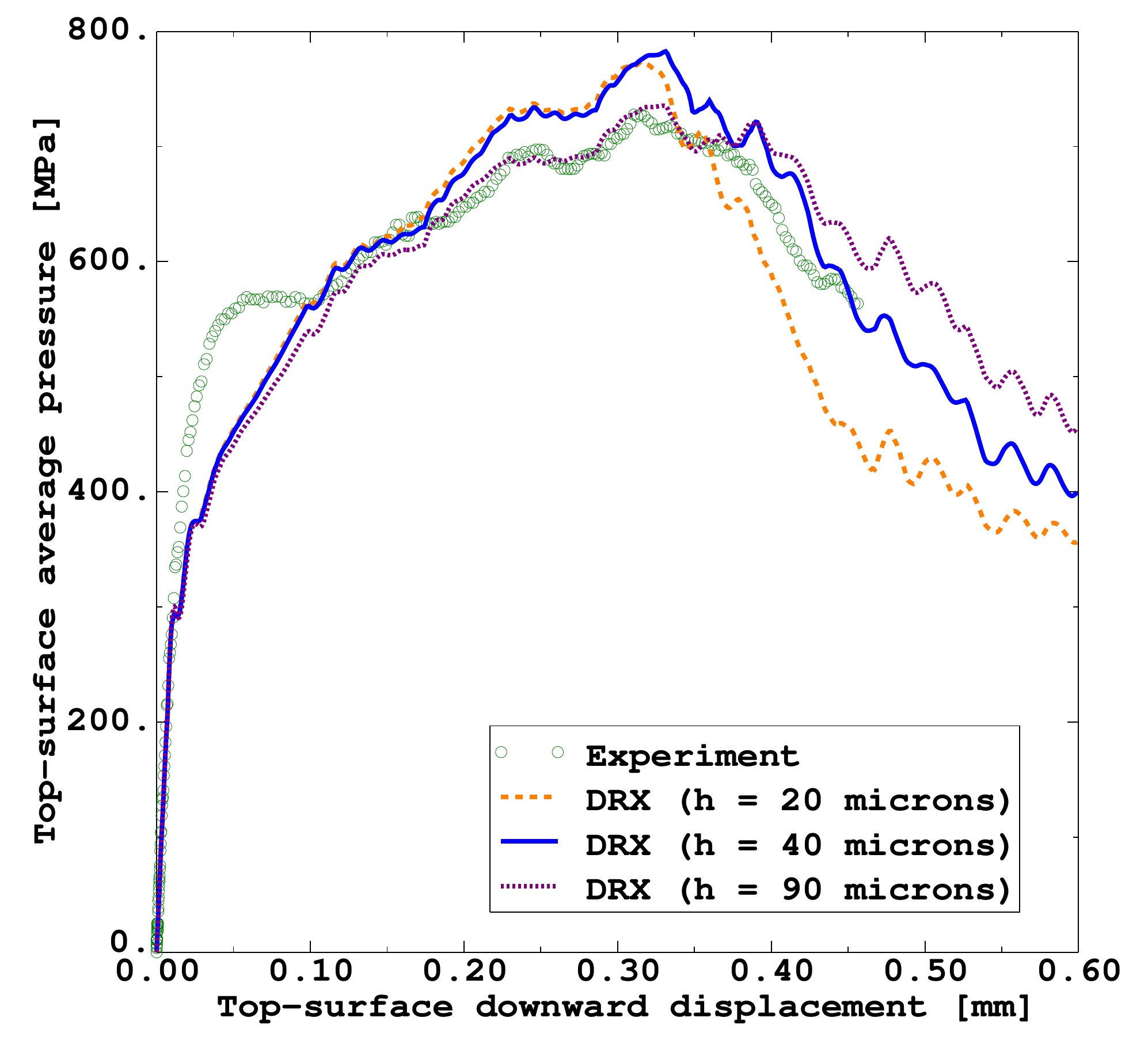}
\caption{\label{fig:splotm}Load-displacement curves computed using the finite element method with $h = 20$, 40, and 90 $\mu$m, compared to experimental measurements.}
\end{center}
\end{figure}

Fig.~\ref{fig:splotm} shows the load-displacement curves computed using the conventional finite element method, and the implicit algorithm described above, for the mesh sizes $h = 20$, 40, and 90 $\mu$m in the shear section. Note that the stress drop increases slightly with decreasing mesh size, which artificially sets the ASB width. This conforms with the intuition that localization of plastic work within a narrower band increases the thermal heating and therefore the thermal softening and recrystallization activity in the band, thereby accounting for the greater stress drop. The mesh-dependence issue can be addressed by embedding the assumed ASB width into the mesh, by means of sub-grid methods~\citep[e.g.,][]{mourad_2017,jin_2018}; this is beyond the scope of the present work. The stress drop for $h = 40 \mu$m appears to be in closet agreement with the experiment; we shall focus on $h = 40\mu$m henceforth in this paper.

\begin{figure}
\begin{center}
\includegraphics[width=0.6\textwidth]{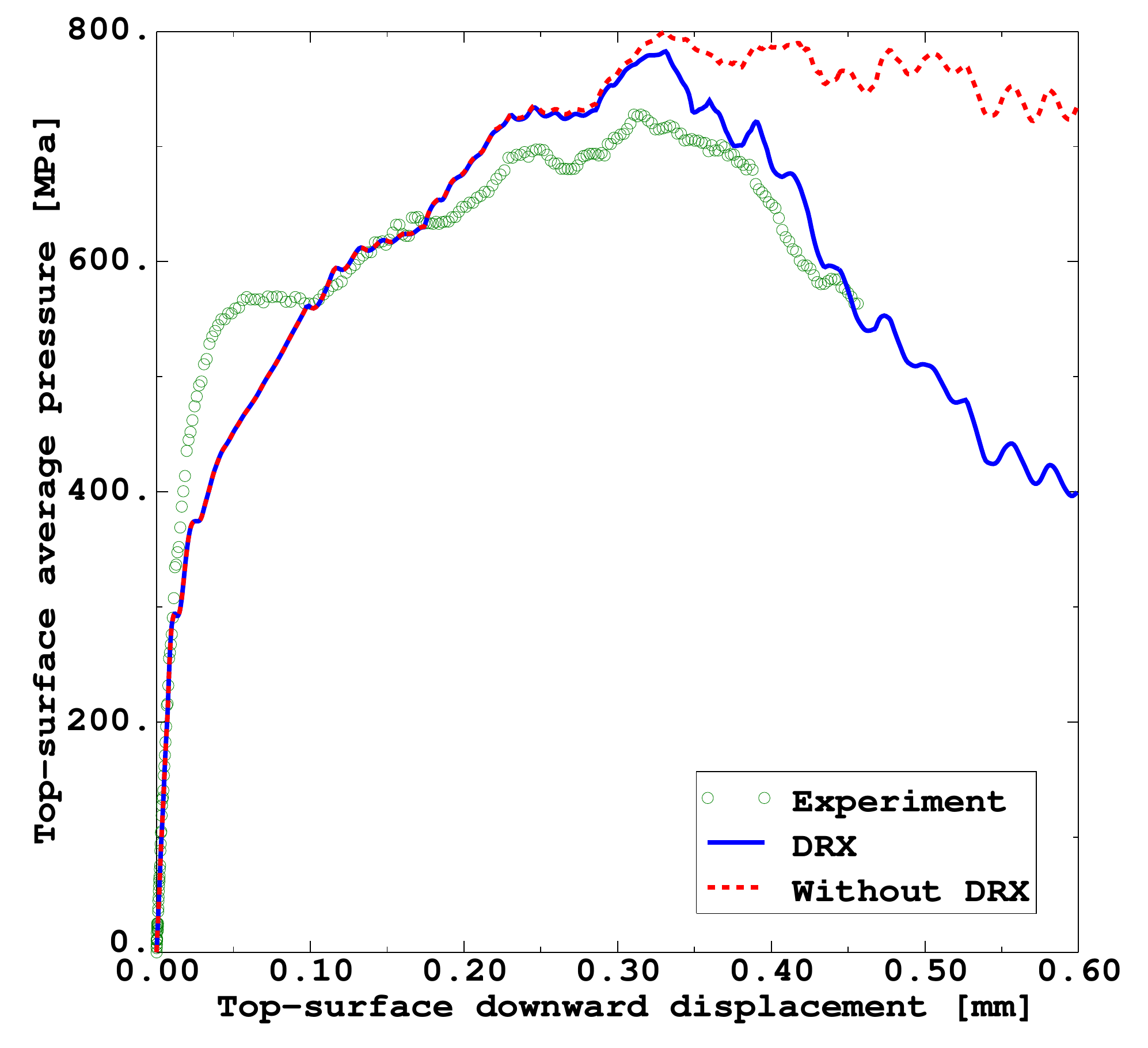}
\caption{\label{fig:splot}Load-displacement curve computed using the finite element method with $h = 40 \mu$m, compared to experimental measurements. Also shown is the load-displacement curve computed for ``pseudo-steel'' that does not undergo dynamic recrystallization, and with otherwise identical material parameters, to indicate the crucial role of DRX in material softening.}
\end{center}
\end{figure}

To demonstrate the softening effect of dynamic recrystallization, we performed the simulation with ``pseudo-steel'', for which DRX is prohibited by setting $\kappa_d = 0$, but whose material parameters are otherwise identical to those listed in Table \ref{tab:parameters} for 316L stainless steel. The resulting load-displacement curve is shown in Fig.~\ref{fig:splot} alongside the result for 316L stainless steel and the experimental measurements; the stress drop upon the formation of the shear band is almost negligible. This result indicates that DRX provides a crucial softening mechanism and may be needed to explain the observed stress drop.

\begin{figure}
\begin{center}
\includegraphics[width=1.0\textwidth]{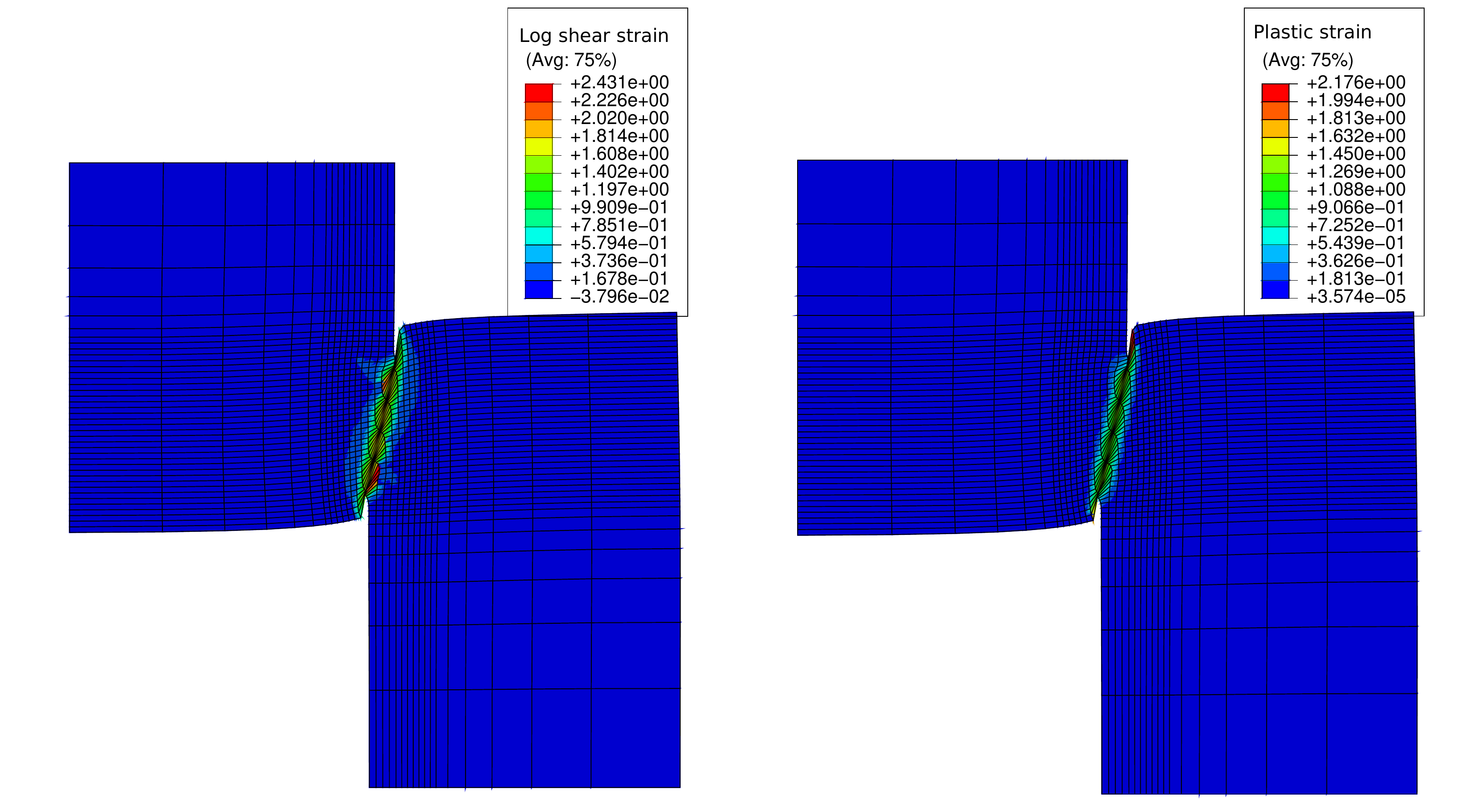}
\caption{\label{fig:eplot}Logarithmic shear strain (left) and plastic strain invariant (right) accumulated at the end of the experiment.}
\end{center}
\end{figure}

\begin{figure}
\begin{center}
\includegraphics[width=1.0\textwidth]{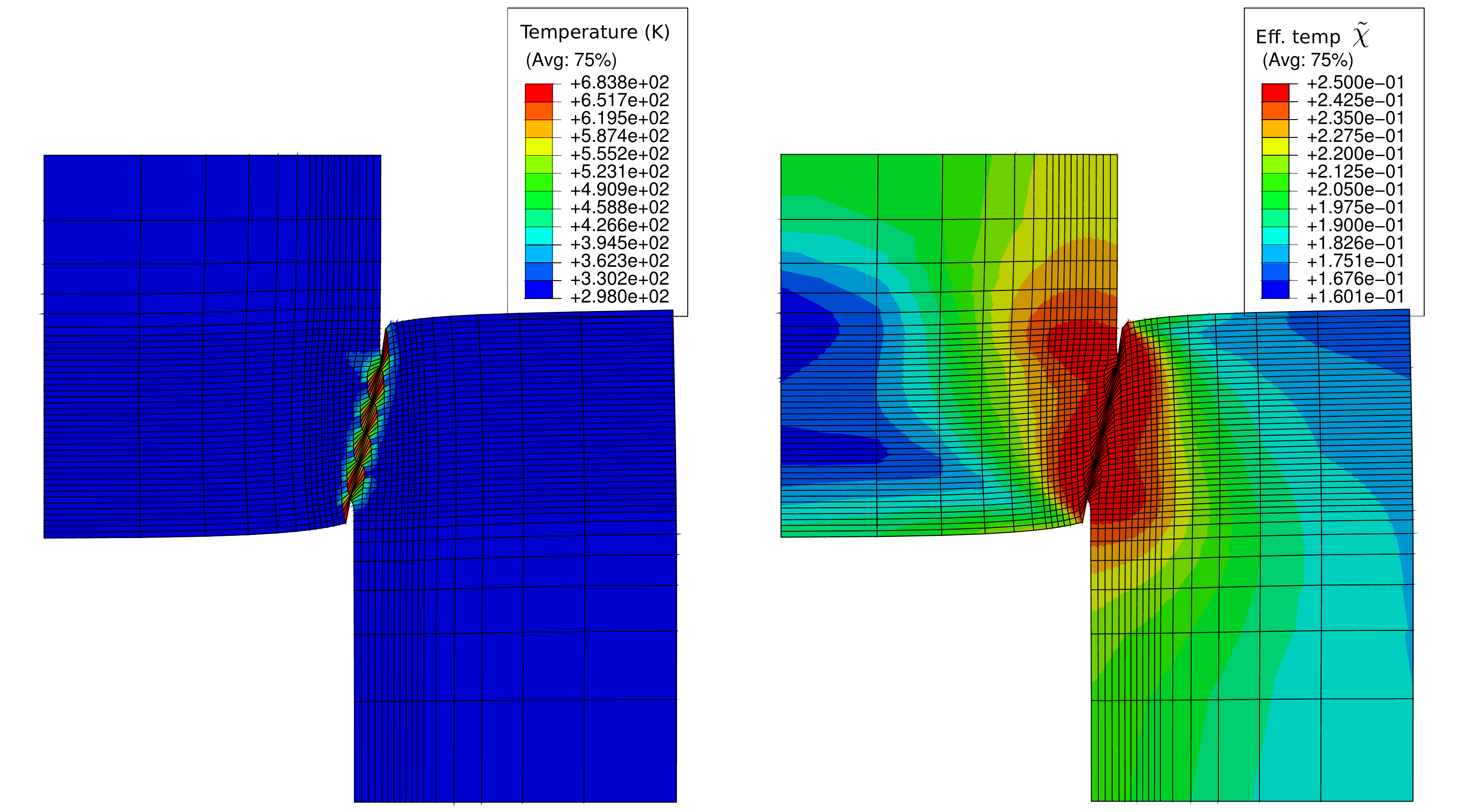}
\caption{\label{fig:Tplot}Model prediction for the distribution of the temperature $T$ (left) and dimensionless effective temperature $\tilde{\chi}$ (right) in the hat-shaped sample, at the end of the experiment. Within the adiabatic assumption, thermal heating occurs almost exclusively within the shear band. The effective temperature saturates to $\tilde{\chi}_0 = 0.25$ in the shear band, but not far away from it.}
\end{center}
\end{figure}

To verify the position of the shear band, we show in Fig.~\ref{fig:eplot} the logarithmic shear strain and accumulated plastic strain $\bar{\epsilon}^{\text{pl}} \equiv \sqrt{(1/2) \epsilon_{ij}^{\text{pl}} \epsilon_{ij}^{\text{pl}}}$, where $\epsilon_{ij}^{\text{pl}} \equiv \int_0^t dt' \, \dot{\epsilon}_{ij}^{\text{pl}} (t')$, at the end of the experiment. Model prediction for the temperature rise in the ASB is given by the left panel Fig.~\ref{fig:Tplot}, which shows the concentration of the heat generated by the plastic work within the shear band. The predicted temperature rise is very close to that given by the MTS model coupled with the sub-grid finite element formulation \citep{mourad_2017}. The right panel of Fig.~\ref{fig:Tplot} shows the distribution of the effective temperature $\tilde{\chi}$; The increase and subsequent saturation of $\tilde{\chi}$ within the shear band, or the growth of configurational disorder therein, causes the evolution of both the dislocation density and the grain size to the $\tilde{\chi}$-controlled values $\tilde{\rho}^{\text{ss}}$ and $d^{\text{ss}}$ given by Eqs.~\eqref{eq:rho_ss2} and \eqref{eq:xi_ss2}. Because $\tilde{\chi}$ increases at a rate proportional to the plastic work, it changes little far away from the ASB.

\begin{figure}
\begin{center}
\begin{subfigure}[t]{\textwidth}
\includegraphics[width=1.0\textwidth]{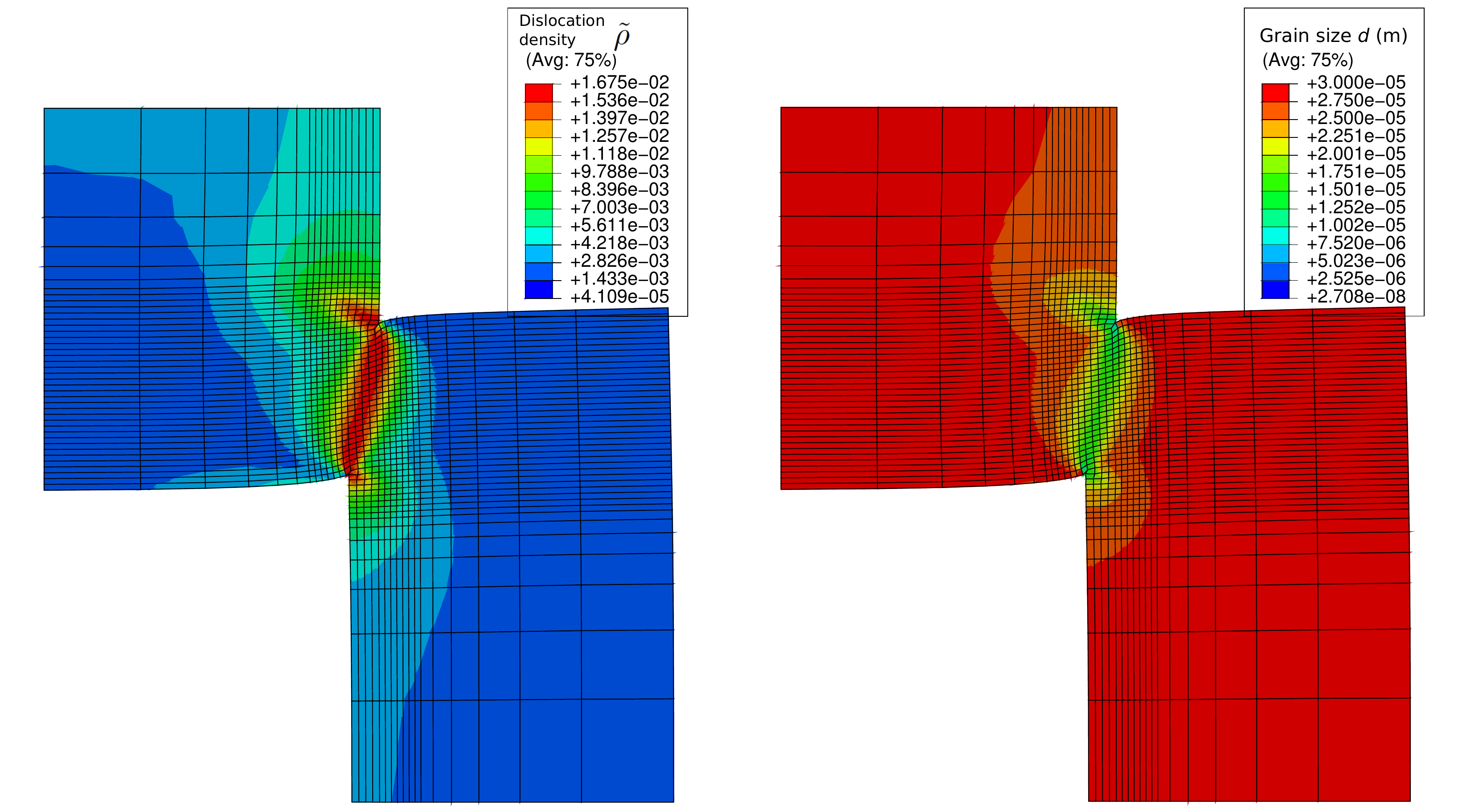}\caption{}
\end{subfigure}
%\vspace{2mm}
\begin{subfigure}[t]{\textwidth}
\includegraphics[width=1.0\textwidth]{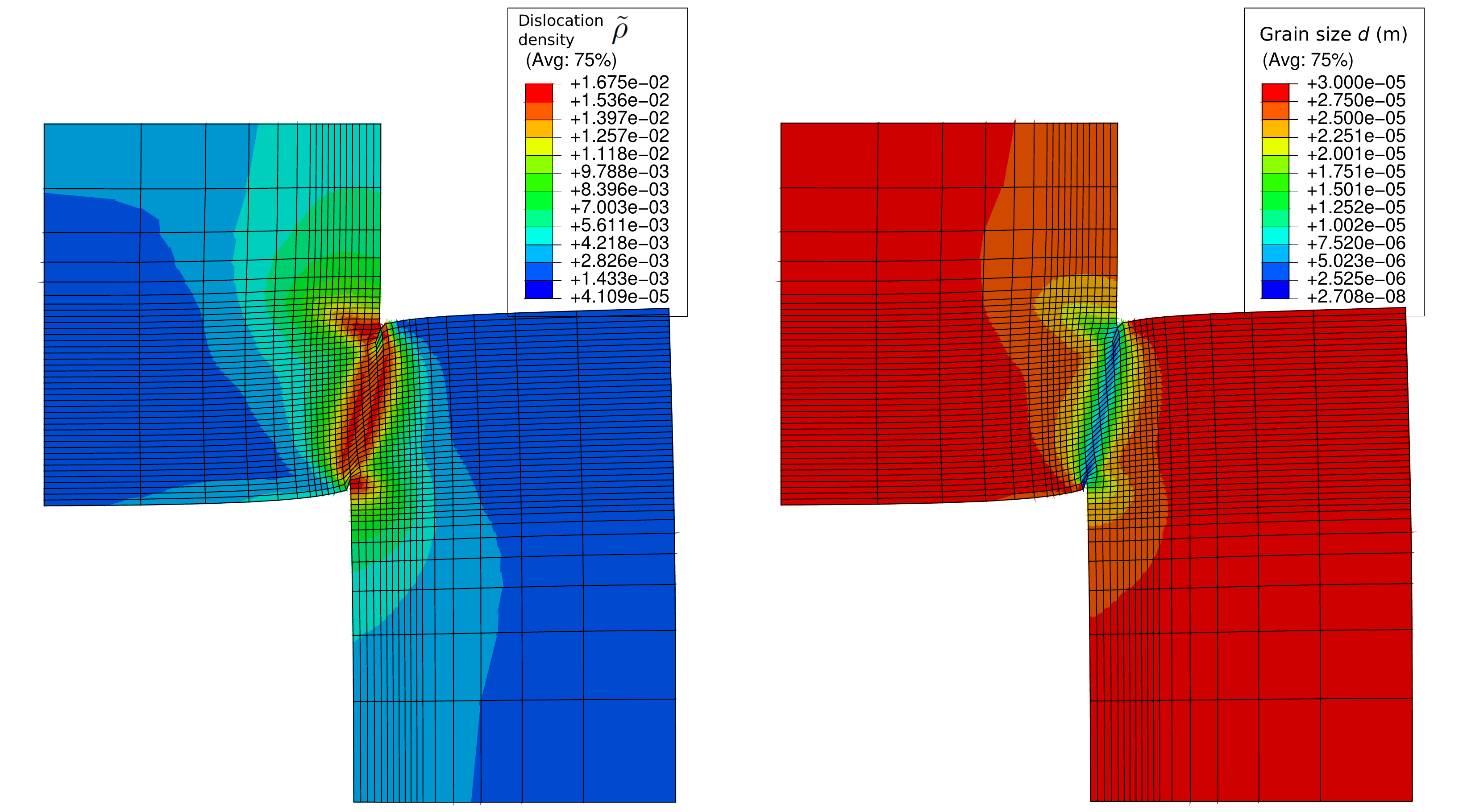}\caption{}
\end{subfigure}
\end{center}
\end{figure}
\begin{figure}\ContinuedFloat
\begin{center}
\begin{subfigure}[t]{\textwidth}
\includegraphics[width=1.0\textwidth]{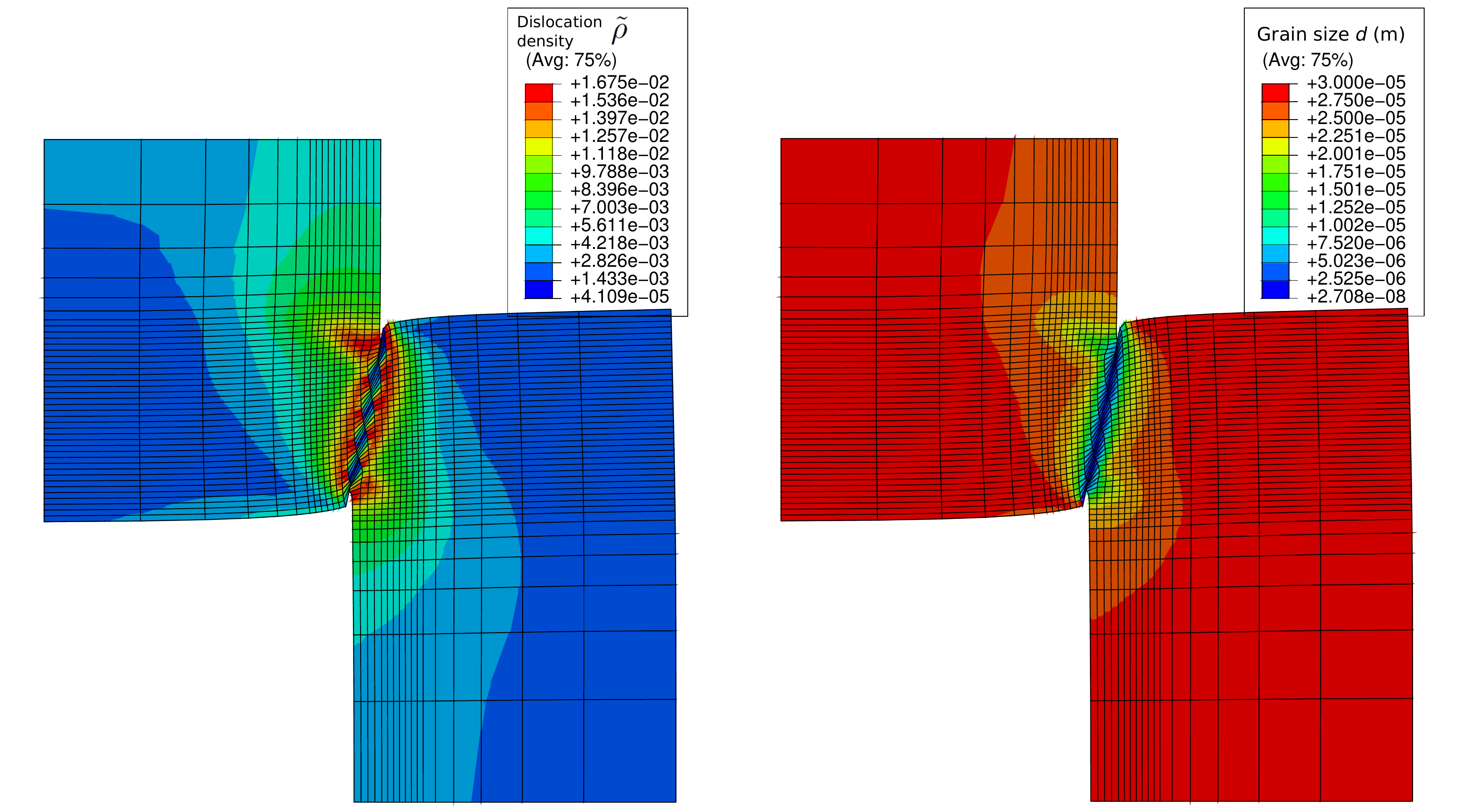}\caption{}
\end{subfigure}
\begin{subfigure}[t]{\textwidth}
\includegraphics[width=1.0\textwidth]{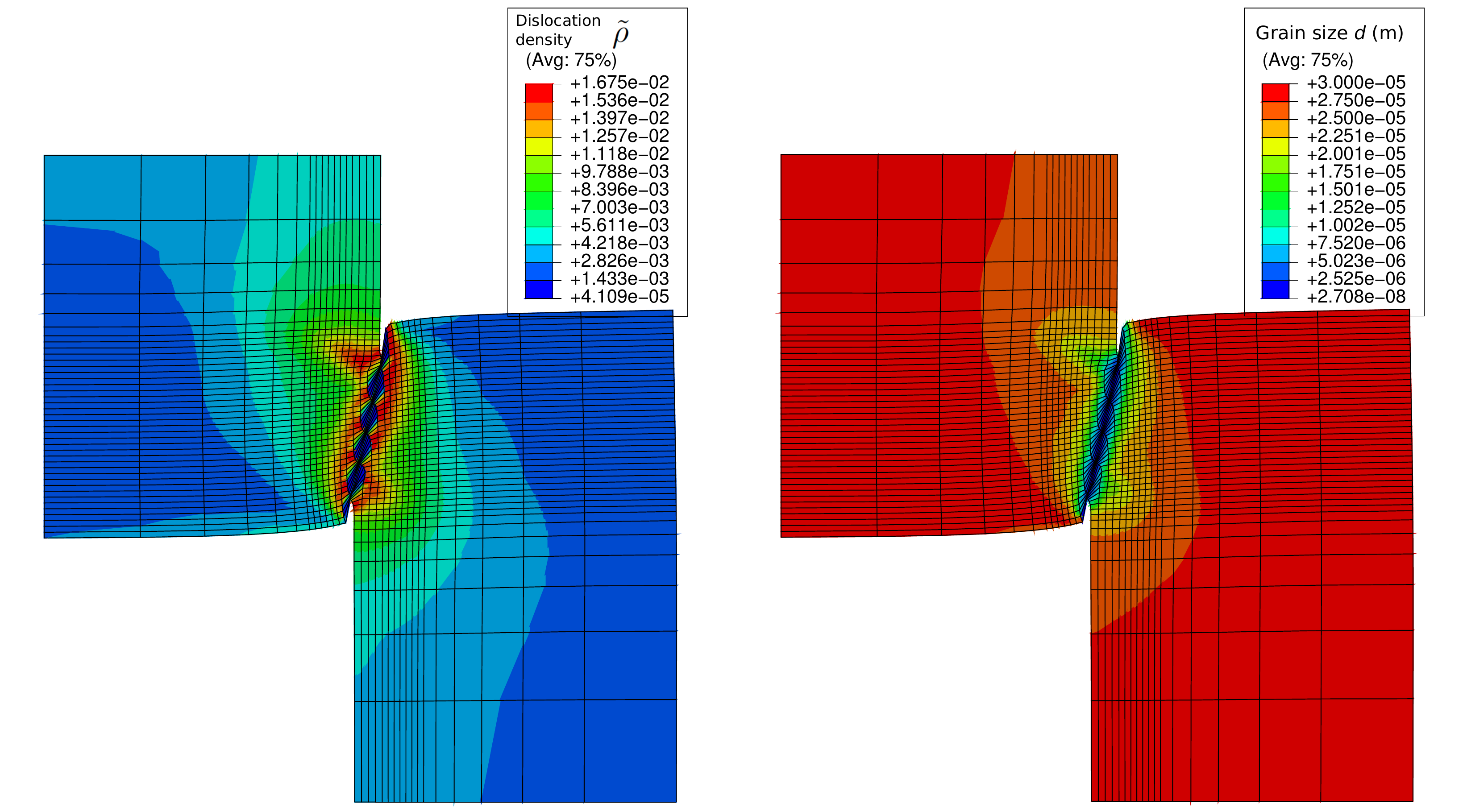}\caption{}
\end{subfigure}
\end{center}
\end{figure}
\begin{figure}\ContinuedFloat
\begin{center}
\begin{subfigure}[t]{\textwidth}
\begin{center}
\includegraphics[width=0.6\textwidth]{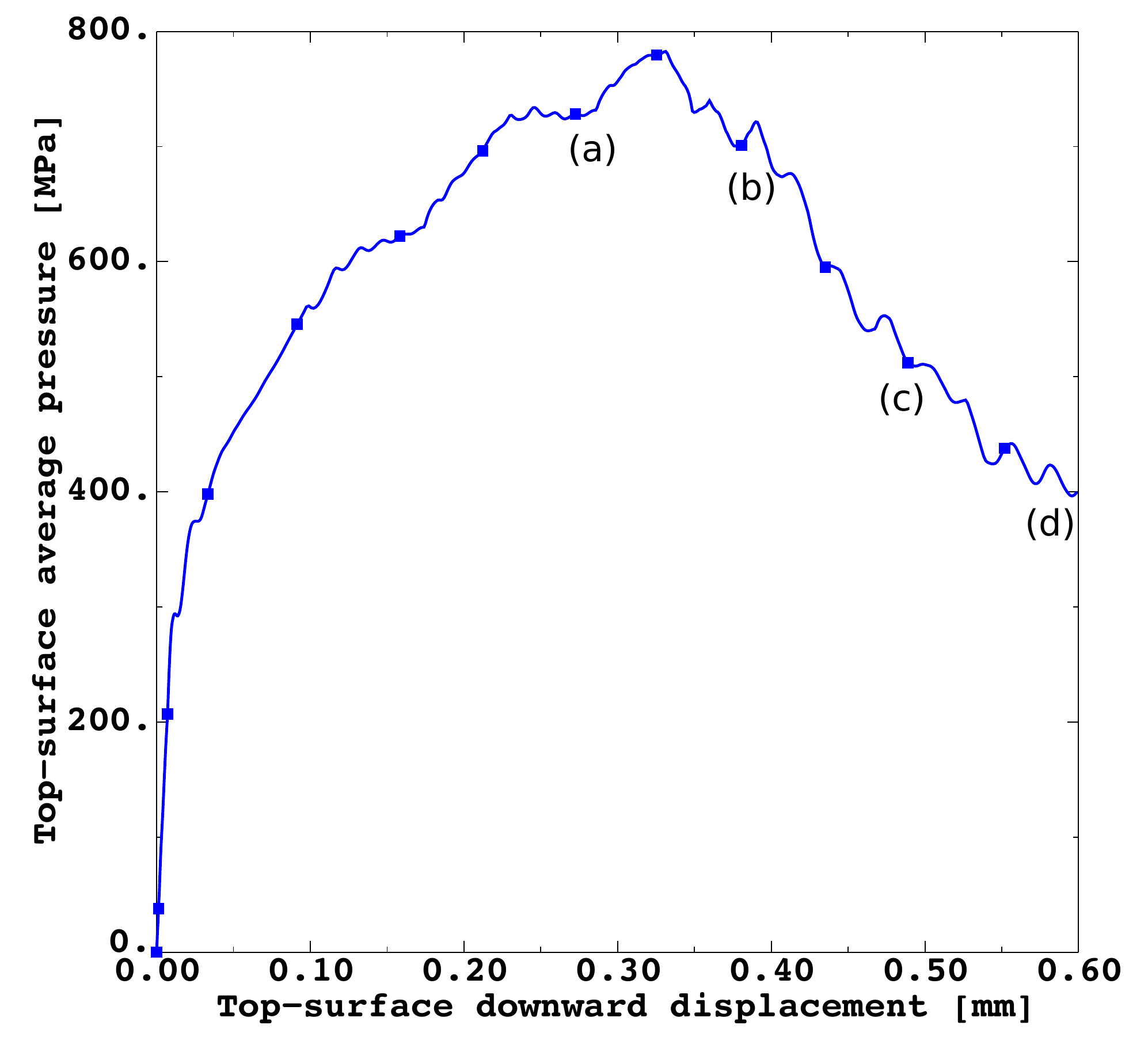}\caption{\label{fig:stress_marked}}
\end{center}
\end{subfigure}
\caption{\label{fig:rhodplot}Evolution of the dimensionless dislocation density $\tilde{\rho}$ (left), and the grain size $d$. Snapshots are taken at the instances marked (a) - (d) in panel (e) at the bottom.}
\end{center}
\end{figure}

Turn now to our predictions for microstructural evolution. Fig.~\ref{fig:rhodplot} shows four snapshots of the dislocation density and grain size profiles in the hat-shaped sample, at the four instances marked (a) through (d) in the bottom panel \ref{fig:stress_marked} in that figure. We see an initial growth of the dislocation density in the shear band, accompanied by a mild decrease of the characteristic grain size, apparently representative of the initial nucleation of DRX grains at grain junctions as described in \citep{takaki_2008,takaki_2009}. When the strain rate in the ASB becomes large enough, dislocations are converted into new grain boundaries. The grain size $d$ within the ASB at the end of the experiment goes down to 270 nm, two orders of magnitude below the initial grain size, while the dislocation density $\tilde{\rho}$ decreases concomitantly by more than two orders of magnitude, to a value even below that of the initial dislocation density. While our quantitative predictions need further verification from more advanced imaging techniques, which will further constrain the parameters that control the rates of grain size reduction and dislocation depletion, these results suggest the possibility of using severe loading conditions to produce ultrafine-grained material almost free of dislocations.

\section{Concluding remarks}
\label{sec:6}

This paper presents the first implementation of the thermodynamic theory of dislocation plasticity and dynamic recrystallization in a finite-element simulation framework. Using known parameters for steel, plus a small handful of tunable parameters with available order-of-magnitude estimates, we have been able to fit the experimental stress-strain behavior of 316L stainless steel that undergoes dynamic recrystallization, using only minimal assumptions. In so doing, we have also provided a simple estimate of the Taylor-Quinney coefficient. This serves as an indication of the validity and usefulness of the LBL-DRX theory. Our effort represents a first step in bridging between physical theories and numerical simulations; there remains substantial work to be done in this area, including more sophisticated finite-element implementations that address the issue of shear-band width and mesh dependency. We conclude with a plea for more detailed imaging and thermomechanical measurements of the recrystallization process, in order to shed light on the nature of ASBs and further validate the theory via a more rigorous constraint of parameters appropriate for steel and other materials.

\section*{Acknowledgements}

CL was partially supported by the Center for Nonlinear Studies at the Los Alamos National Laboratory over the duration of this work. All authors were partially supported by the DOE/DOD Joint Munitions Program and and LANL LDRD Program Project 20170033DR. The authors declare no conflicts of interest. 

%% The Appendices part is started with the command \appendix;
%% appendix sections are then done as normal sections
%% \appendix

%% \section{}
%% \label{}

%% If you have bibdatabase file and want bibtex to generate the
%% bibitems, please use
%%
\bibliographystyle{elsarticle-harv} 
\bibliography{ijp_drxsteel_03}

\begin{thebibliography}{66}
\expandafter\ifx\csname natexlab\endcsname\relax\def\natexlab#1{#1}\fi
\expandafter\ifx\csname url\endcsname\relax
  \def\url#1{\texttt{#1}}\fi
\expandafter\ifx\csname urlprefix\endcsname\relax\def\urlprefix{URL }\fi

\bibitem[{Abed and Voyiadjis(2005)}]{abed_2005}
Abed, F.~H., Voyiadjis, G.~Z., 2005. Plastic deformation modeling of al-6xn
  stainless steel at low and high strain rates and temperatures using a
  combination of bcc and fcc mechanisms of metals. International Journal of
  Plasticity 21~(8), 1618--1639.
\newline\urlprefix\url{http://www.sciencedirect.com/science/article/pii/S0749641904001652}

\bibitem[{Abed and Voyiadjis(2007)}]{abed_2007}
Abed, F.~H., Voyiadjis, G.~Z., 2007. Adiabatic shear band localizations in bcc
  metals at high strain rates and various initial temperatures. International
  Journal for Multiscale Computational Engineering 5~(3-4), 325--349.

\bibitem[{Ahad et~al.(2014)Ahad, Enakoutsa, Solanski, and Bammann}]{ahad_2014}
Ahad, F.~R., Enakoutsa, K., Solanski, K.~N., Bammann, D.~J., 2014. Nonlocal
  modeling in high-velocity impact failure of 6061-t6 aluminum. International
  Journal of Plasticity 55, 108--132.
\newline\urlprefix\url{http://www.sciencedirect.com/science/article/pii/S0749641913001927}

\bibitem[{Anand et~al.(2012)Anand, Aslan, and Chester}]{anand_2012}
Anand, L., Aslan, O., Chester, S.~A., 2012. A large-deformation gradient theory
  for elastic-plastic materials: Strain softening and regularization of shear
  bands. International Journal of Plasticity 30-31, 115--143.
\newline\urlprefix\url{http://www.sciencedirect.com/science/article/pii/S0749641911001665}

\bibitem[{Anand et~al.(2015)Anand, Gurtin, and Reddy}]{anand_2015}
Anand, L., Gurtin, M.~E., Reddy, B.~D., 2015. The stored energy of cold work,
  thermal annealing, and other thermodynamic issues in single crystal
  plasticity at small length scales. International Journal of Plasticity 64, 1
  -- 25.
\newline\urlprefix\url{http://www.sciencedirect.com/science/article/pii/S074964191400148X}

\bibitem[{Arriaga and Waisman(2017)}]{arriaga_2017}
Arriaga, M., Waisman, H., 2017. Combined stability analysis of phase-field
  dynamic fracture and shear band localization. International Journal of
  Plasticity 96, 81 -- 119.
\newline\urlprefix\url{http://www.sciencedirect.com/science/article/pii/S0749641916303175}

\bibitem[{Benzerga et~al.(2005)Benzerga, Bréchet, Needleman, and der
  Giessen}]{benzerga_2005}
Benzerga, A., Bréchet, Y., Needleman, A., der Giessen, E.~V., 2005. The stored
  energy of cold work: Predictions from discrete dislocation plasticity. Acta
  Materialia 53~(18), 4765 -- 4779.
\newline\urlprefix\url{http://www.sciencedirect.com/science/article/pii/S1359645405003964}

\bibitem[{Bouchbinder and Langer(2009)}]{bouchbinder_2009b}
Bouchbinder, E., Langer, J.~S., Sep 2009. Nonequilibrium thermodynamics of
  driven amorphous materials. ii. effective-temperature theory. Phys. Rev. E
  80, 031132.
\newline\urlprefix\url{https://link.aps.org/doi/10.1103/PhysRevE.80.031132}

\bibitem[{Bronkhorst et~al.(2006)Bronkhorst, Cerreta, Xue, Maudlin, Mason, and
  Gray}]{bronkhorst_2006}
Bronkhorst, C., Cerreta, E., Xue, Q., Maudlin, P., Mason, T., Gray, G., 2006.
  An experimental and numerical study of the localization behavior of tantalum
  and stainless steel. International Journal of Plasticity 22~(7), 1304 --
  1335.
\newline\urlprefix\url{http://www.sciencedirect.com/science/article/pii/S0749641905001592}

\bibitem[{Brown and Bammann(2012)}]{brown_2012}
Brown, A.~A., Bammann, D.~J., 2012. Validation of a model for static and
  dynamic recrystallization in metals. International Journal of Plasticity
  32-33, 17--35.
\newline\urlprefix\url{http://www.sciencedirect.com/science/article/pii/S0749641911001963}

\bibitem[{Coleman and Noll(1963)}]{coleman_1963}
Coleman, B.~D., Noll, W., Dec 1963. The thermodynamics of elastic materials
  with heat conduction and viscosity. Archive for Rational Mechanics and
  Analysis 13~(1), 167--178.
\newline\urlprefix\url{https://doi.org/10.1007/BF01262690}

\bibitem[{Dodd and Bai(2012)}]{dodd_2012}
Dodd, B., Bai, Y. (Eds.), 2012. Adiabatic Shear Localization, Frontiers and
  Advances, 2nd Edition. Elsevier, Oxford.

\bibitem[{Duffy and Chi(1992)}]{duffy_1992}
Duffy, J., Chi, Y., 1992. On the measurement of local strain and temperature
  during the formation of adiabatic shear bands. Materials Science and
  Engineering: A 157~(2), 195 -- 210.
\newline\urlprefix\url{http://www.sciencedirect.com/science/article/pii/092150939290026W}

\bibitem[{Farren and Taylor(1925)}]{farren_1925}
Farren, W.~S., Taylor, G.~I., 1925. The heat developed during plastic extension
  of metals. Proceedings of the Royal Society of London A: Mathematical,
  Physical and Engineering Sciences 107~(743), 422--451.
\newline\urlprefix\url{http://rspa.royalsocietypublishing.org/content/107/743/422}

\bibitem[{Galindo-Nava and del Castillo(2014)}]{galindo-nava_2014}
Galindo-Nava, E., del Castillo, P. R.-D., 2014. Grain size evolution during
  discontinuous dynamic recrystallization. Scripta Materialia 72-73, 1 -- 4.
\newline\urlprefix\url{http://www.sciencedirect.com/science/article/pii/S1359646213004764}

\bibitem[{Galindo-Nava and Rae(2015)}]{galindo-nava_2015}
Galindo-Nava, E., Rae, C., 2015. Microstructure evolution during dynamic
  recrystallisation in polycrystalline nickel superalloys. Materials Science
  and Engineering: A 636, 434 -- 445.
\newline\urlprefix\url{http://www.sciencedirect.com/science/article/pii/S0921509315003810}

\bibitem[{Hartley et~al.(1987)Hartley, Duffy, and Hawley}]{hartley_1987}
Hartley, K., Duffy, J., Hawley, R., 1987. Measurement of the temperature
  profile during shear band formation in steels deforming at high strain rates.
  Journal of the Mechanics and Physics of Solids 35~(3), 283 -- 301.
\newline\urlprefix\url{http://www.sciencedirect.com/science/article/pii/0022509687900093}

\bibitem[{Hines and Vecchio(1997)}]{hines_1997}
Hines, J., Vecchio, K., 1997. Recrystallization kinetics within adiabatic shear
  bands. Acta Materialia 45~(2), 635 -- 649.
\newline\urlprefix\url{http://www.sciencedirect.com/science/article/pii/S1359645496001930}

\bibitem[{Hines et~al.(1998)Hines, Vecchio, and Ahzi}]{hines_1998}
Hines, J.~A., Vecchio, K.~S., Ahzi, S., Jan 1998. A model for microstructure
  evolution in adiabatic shear bands. Metallurgical and Materials Transactions
  A 29~(1), 191--203.
\newline\urlprefix\url{https://doi.org/10.1007/s11661-998-0172-4}

\bibitem[{Jin et~al.(2018)Jin, Mourad, Bronkhorst, and Livescu}]{jin_2018}
Jin, T., Mourad, H.~M., Bronkhorst, C.~A., Livescu, V., Feb 2018. Finite
  element formulation with embedded weak discontinuities for strain
  localization under dynamic conditions. Computational Mechanics 61~(1), 3--18.
\newline\urlprefix\url{https://doi.org/10.1007/s00466-017-1470-8}

\bibitem[{Kocks(1966)}]{kocks_1966}
Kocks, U.~F., 1966. A statistical theory of flow stress and work-hardening. The
  Philosophical Magazine: A Journal of Theoretical Experimental and Applied
  Physics 13~(123), 541--566.
\newline\urlprefix\url{https://doi.org/10.1080/14786436608212647}

\bibitem[{Langer et~al.(2010)Langer, Bouchbinder, and Lookman}]{langer_2010}
Langer, J., Bouchbinder, E., Lookman, T., 2010. Thermodynamic theory of
  dislocation-mediated plasticity. Acta Materialia 58~(10), 3718 -- 3732.
\newline\urlprefix\url{http://www.sciencedirect.com/science/article/pii/S1359645410001540}

\bibitem[{Langer(2015)}]{langer_2015}
Langer, J.~S., Sep 2015. Statistical thermodynamics of strain hardening in
  polycrystalline solids. Phys. Rev. E 92, 032125.
\newline\urlprefix\url{https://link.aps.org/doi/10.1103/PhysRevE.92.032125}

\bibitem[{Langer(2016)}]{langer_2016}
Langer, J.~S., Dec 2016. Thermal effects in dislocation theory. Phys. Rev. E
  94, 063004.
\newline\urlprefix\url{https://link.aps.org/doi/10.1103/PhysRevE.94.063004}

\bibitem[{Langer(2017{\natexlab{a}})}]{langer_2017a}
Langer, J.~S., Jan 2017{\natexlab{a}}. Thermal effects in dislocation theory.
  ii. shear banding. Phys. Rev. E 95, 013004.
\newline\urlprefix\url{https://link.aps.org/doi/10.1103/PhysRevE.95.013004}

\bibitem[{Langer(2017{\natexlab{b}})}]{langer_2017b}
Langer, J.~S., Mar 2017{\natexlab{b}}. Yielding transitions and grain-size
  effects in dislocation theory. Phys. Rev. E 95, 033004.
\newline\urlprefix\url{https://link.aps.org/doi/10.1103/PhysRevE.95.033004}

\bibitem[{Le et~al.(2018)Le, Tran, and Langer}]{le_2018}
Le, K., Tran, T., Langer, J., 2018. Thermodynamic dislocation theory of
  adiabatic shear banding in steel. Scripta Materialia 149, 62 -- 65.
\newline\urlprefix\url{http://www.sciencedirect.com/science/article/pii/S1359646218300836}

\bibitem[{Li et~al.(2017)Li, Wang, Zhao, Valiev, Vecchio, and Meyers}]{li_2017}
Li, Z., Wang, B., Zhao, S., Valiev, R.~Z., Vecchio, K.~S., Meyers, M.~A., 2017.
  Dynamic deformation and failure of ultrafine-grained titanium. Acta
  Materialia 125~(Supplement C), 210 -- 218.
\newline\urlprefix\url{http://www.sciencedirect.com/science/article/pii/S1359645416309089}

\bibitem[{Lieou and Bronkhorst(2018)}]{lieou_2018}
Lieou, C.~K., Bronkhorst, C.~A., 2018. Dynamic recrystallization in adiabatic
  shear banding: Effective-temperature model and comparison to experiments in
  ultrafine-grained titanium. International Journal of Plasticity.
\newline\urlprefix\url{http://www.sciencedirect.com/science/article/pii/S0749641918302821}

\bibitem[{Lindemann(1910)}]{lindemann_1910}
Lindemann, F.~A., 1910. The calculation of molecular vibration frequencies.
  Physik. Z. 11, 609--612.

\bibitem[{Longère and Dragon(2008{\natexlab{a}})}]{longere_2008b}
Longère, P., Dragon, A., 2008{\natexlab{a}}. Evaluation of the inelastic heat
  fraction in the context of microstructure-supported dynamic plasticity
  modelling. International Journal of Impact Engineering 35~(9), 992 -- 999.
\newline\urlprefix\url{http://www.sciencedirect.com/science/article/pii/S0734743X07001054}

\bibitem[{Longère and Dragon(2008{\natexlab{b}})}]{longere_2008a}
Longère, P., Dragon, A., 2008{\natexlab{b}}. Plastic work induced heating
  evaluation under dynamic conditions: Critical assessment. Mechanics Research
  Communications 35~(3), 135 -- 141.
\newline\urlprefix\url{http://www.sciencedirect.com/science/article/pii/S0093641307000894}

\bibitem[{Luscher et~al.(2018)Luscher, Buechler, Walters, Bolme, and
  Ramos}]{luscher_2018}
Luscher, D.~J., Buechler, M.~A., Walters, D.~J., Bolme, C., Ramos, K.~J., 2018.
  On computing the evolution of temperature for materials under dynamic
  loading. International Journal of Plasticity.
\newline\urlprefix\url{http://www.sciencedirect.com/science/article/pii/S0749641918301426}

\bibitem[{Marchand and Duffy(1988)}]{marchand_1988}
Marchand, A., Duffy, J., 1988. An experimental study of the formation process
  of adiabatic shear bands in a structural steel. Journal of the Mechanics and
  Physics of Solids 36~(3), 251 -- 283.
\newline\urlprefix\url{http://www.sciencedirect.com/science/article/pii/0022509688900129}

\bibitem[{Mecking and Kocks(1981)}]{mecking_1981}
Mecking, H., Kocks, U., 1981. Kinetics of flow and strain-hardening. Acta
  Metallurgica 29~(11), 1865 -- 1875.
\newline\urlprefix\url{http://www.sciencedirect.com/science/article/pii/0001616081901127}

\bibitem[{Meyer and Manwaring(1985)}]{meyer_1985}
Meyer, L.~W., Manwaring, S., 1985. Critical adiabatic shear strength of low
  alloyed steel under compressive loading. In: International Conference on
  Metallurgical Applications of Shock-Wave and High-Strain-Rate
  Phenomena(EXPLOMET85). pp. 657--674.

\bibitem[{Meyers et~al.(2003)Meyers, Xu, Xue, Pérez-Prado, and
  McNelley}]{meyers_2003}
Meyers, M., Xu, Y., Xue, Q., Pérez-Prado, M., McNelley, T., 2003.
  Microstructural evolution in adiabatic shear localization in stainless steel.
  Acta Materialia 51~(5), 1307 -- 1325.
\newline\urlprefix\url{http://www.sciencedirect.com/science/article/pii/S1359645402005268}

\bibitem[{Meyers et~al.(2000)Meyers, Nesterenko, LaSalvia, Xu, and
  Xue}]{meyers_2000}
Meyers, M.~A., Nesterenko, V.~F., LaSalvia, J.~C., Xu, Y.~B., Xue, Q., 2000.
  Observation and modeling of dynamic recrystallization in high-strain,
  high-strain rate deformation of metals. Journal of Physics IV France 10,
  51--56.
\newline\urlprefix\url{https://doi.org/10.1051/jp4:2000909}

\bibitem[{Meyers et~al.(2001)Meyers, Nesterenko, LaSalvia, and
  Xue}]{meyers_2001}
Meyers, M.~A., Nesterenko, V.~F., LaSalvia, J.~C., Xue, Q., 2001. Shear
  localization in dynamic deformation of materials: microstructural evolution
  and self-organization. Materials Science and Engineering: A 317~(1),
  204--225.
\newline\urlprefix\url{http://www.sciencedirect.com/science/article/pii/S0921509301011601}

\bibitem[{Mourad et~al.(2017)Mourad, Bronkhorst, Livescu, Plohr, and
  Cerreta}]{mourad_2017}
Mourad, H., Bronkhorst, C., Livescu, V., Plohr, J., Cerreta, E., 2017. Modeling
  and simulation framework for dynamic strain localization in
  elasto-viscoplastic metallic materials subject to large deformations.
  International Journal of Plasticity 88~(Supplement C), 1 -- 26.
\newline\urlprefix\url{http://www.sciencedirect.com/science/article/pii/S0749641916301668}

\bibitem[{Osovski et~al.(2012)Osovski, Rittel, Landau, and
  Venkert}]{osovski_2012}
Osovski, S., Rittel, D., Landau, P., Venkert, A., 2012. Microstructural effects
  on adiabatic shear band formation. Scripta Materialia 66~(1), 9 -- 12.
\newline\urlprefix\url{http://www.sciencedirect.com/science/article/pii/S1359646211005409}

\bibitem[{Osovski et~al.(2013)Osovski, Rittel, and Venkert}]{osovski_2013}
Osovski, S., Rittel, D., Venkert, A., 2013. The respective influence of
  microstructural and thermal softening on adiabatic shear localization.
  Mechanics of Materials 56~(Supplement C), 11 -- 22.
\newline\urlprefix\url{http://www.sciencedirect.com/science/article/pii/S0167663612001664}

\bibitem[{Popova et~al.(2015)Popova, Staraselski, Brahme, Mishra, and
  Inal}]{popova_2015}
Popova, E., Staraselski, Y., Brahme, A., Mishra, R., Inal, K., 2015. Coupled
  crystal plasticity – probabilistic cellular automata approach to model
  dynamic recrystallization in magnesium alloys. International Journal of
  Plasticity 66, 85 -- 102, plasticity of Textured Polycrystals In Honor of
  Prof. Paul Van Houtte.
\newline\urlprefix\url{http://www.sciencedirect.com/science/article/pii/S0749641914000916}

\bibitem[{Puchi-Cabrera et~al.(2018)Puchi-Cabrera, Guérin, Barbera-Sosa,
  Dubar, and Dubar}]{puchi-cabrera_2018}
Puchi-Cabrera, E., Guérin, J., Barbera-Sosa, J.~L., Dubar, M., Dubar, L.,
  2018. Plausible extension of anand's model to metals exhibiting dynamic
  recrystallization and its experimental validation. International Journal of
  Plasticity.
\newline\urlprefix\url{http://www.sciencedirect.com/science/article/pii/S0749641917306411}

\bibitem[{Rittel et~al.(2008)Rittel, Landau, and Venkert}]{rittel_2008}
Rittel, D., Landau, P., Venkert, A., Oct 2008. Dynamic recrystallization as a
  potential cause for adiabatic shear failure. Phys. Rev. Lett. 101, 165501.
\newline\urlprefix\url{https://link.aps.org/doi/10.1103/PhysRevLett.101.165501}

\bibitem[{Rittel et~al.(2002)Rittel, Ravichandran, and Lee}]{rittel_2002}
Rittel, D., Ravichandran, G., Lee, S., 2002. Large strain constituve behavior
  of ofhc copper over a wide range of strain rates using the shear compression
  specimen. Mechanics of Materials 34~(10), 627--642.
\newline\urlprefix\url{http://www.sciencedirect.com/science/article/pii/S0167663602001643}

\bibitem[{Rittel et~al.(2006)Rittel, Wang, and Merzer}]{rittel_2006}
Rittel, D., Wang, Z.~G., Merzer, M., Feb 2006. Adiabatic shear failure and
  dynamic stored energy of cold work. Phys. Rev. Lett. 96, 075502.
\newline\urlprefix\url{https://link.aps.org/doi/10.1103/PhysRevLett.96.075502}

\bibitem[{Rittel et~al.(2017)Rittel, Zhang, and Osovski}]{rittel_2017}
Rittel, D., Zhang, L., Osovski, S., 2017. The dependence of the
  taylor–quinney coefficient on the dynamic loading mode. Journal of the
  Mechanics and Physics of Solids 107, 96 -- 114.
\newline\urlprefix\url{http://www.sciencedirect.com/science/article/pii/S0022509617301709}

\bibitem[{Rosakis et~al.(2000)Rosakis, Rosakis, Ravichandran, and
  Hodowany}]{rosakis_2000}
Rosakis, P., Rosakis, A., Ravichandran, G., Hodowany, J., 2000. A thermodynamic
  internal variable model for the partition of plastic work into heat and
  stored energy in metals. Journal of the Mechanics and Physics of Solids
  48~(3), 581 -- 607.
\newline\urlprefix\url{http://www.sciencedirect.com/science/article/pii/S0022509699000484}

\bibitem[{Rousselier and Quilici(2015)}]{rousselier_2015}
Rousselier, G., Quilici, S., 2015. Combining porous plasticity with coulomb and
  portevin-le chatelier models for ductile fracture analyses. International
  Journal of Plasticity 69, 118 -- 133.
\newline\urlprefix\url{http://www.sciencedirect.com/science/article/pii/S0749641915000388}

\bibitem[{Sabnis et~al.(2012)Sabnis, Mazière, Forest, Arakere, and
  Ebrahimi}]{sabnis_2012}
Sabnis, P., Mazière, M., Forest, S., Arakere, N.~K., Ebrahimi, F., 2012.
  Effect of secondary orientation on notch-tip plasticity in superalloy single
  crystals. International Journal of Plasticity 28~(1), 102 -- 123.
\newline\urlprefix\url{http://www.sciencedirect.com/science/article/pii/S0749641911001045}

\bibitem[{Smith(2014)}]{abaqus_2014}
Smith, M., 2014. ABAQUS/Standard User's Manual, Version 6.14. Simulia.

\bibitem[{Stainier and Ortiz(2010)}]{stainier_2010}
Stainier, L., Ortiz, M., 2010. Study and validation of a variational theory of
  thermo-mechanical coupling in finite visco-plasticity. International Journal
  of Solids and Structures 47~(5), 705 -- 715.
\newline\urlprefix\url{http://www.sciencedirect.com/science/article/pii/S0020768309004478}

\bibitem[{Sun et~al.(2018)Sun, Wu, Cao, and Yin}]{sun_2018}
Sun, Z., Wu, H., Cao, J., Yin, Z., 2018. Modeling of continuous dynamic
  recrystallization of al-zn-cu-mg alloy during hot deformation based on the
  internal-state-variable (isv) method. International Journal of Plasticity.
\newline\urlprefix\url{http://www.sciencedirect.com/science/article/pii/S0749641917306381}

\bibitem[{Takaki et~al.(2008)Takaki, Hirouchi, Hisakuni, Yamanaka, and
  Tomita}]{takaki_2008}
Takaki, T., Hirouchi, T., Hisakuni, Y., Yamanaka, A., Tomita, Y., 2008.
  Multi-phase-field model to simulate microstructure evolutions during dynamic
  recrystallization. MATERIALS TRANSACTIONS 49~(11), 2559--2565.

\bibitem[{Takaki et~al.(2009)Takaki, Hisakuni, Hirouchi, Yamanaka, and
  Tomita}]{takaki_2009}
Takaki, T., Hisakuni, Y., Hirouchi, T., Yamanaka, A., Tomita, Y., 2009.
  Multi-phase-field simulations for dynamic recrystallization. Computational
  Materials Science 45~(4), 881 -- 888.
\newline\urlprefix\url{http://www.sciencedirect.com/science/article/pii/S0927025608005247}

\bibitem[{Taylor and Quinney(1934)}]{taylor_1934}
Taylor, G.~I., Quinney, H., 1934. The latent energy remaining in a metal after
  cold working. Proceedings of the Royal Society of London A: Mathematical,
  Physical and Engineering Sciences 143~(849), 307--326.
\newline\urlprefix\url{http://rspa.royalsocietypublishing.org/content/143/849/307}

\bibitem[{Voyiadjis and Abed(2005)}]{voyiadjis_2005}
Voyiadjis, G.~Z., Abed, F.~H., 2005. Microstructural based models for bcc and
  fcc metals with temperature and strain rate dependency. Mechanics of
  Materials 37~(2), 355--378.
\newline\urlprefix\url{http://www.sciencedirect.com/science/article/pii/S0167663604000894}

\bibitem[{Voyiadjis and Abed(2007)}]{voyiadjis_2007}
Voyiadjis, G.~Z., Abed, F.~H., 2007. Transient localizations in metals using
  microstructure-based yield surfaces. Modelling and Simulation in Materials
  Science and Engineering 15~(1), S83--S95.

\bibitem[{Voyiadjis et~al.(2004)Voyiadjis, Abu Al-Rub, and
  Palazotto}]{voyiadjis_2004}
Voyiadjis, G.~Z., Abu Al-Rub, R., Palazotto, A.~N., 2004. Thermodynamic
  framework for coupling of non-local viscoplasticity and non-local anisotropic
  viscodamage for dynamic localization problems using gradient theory.
  International Journal of Plasticity 20~(6), 981--1038.
\newline\urlprefix\url{http://www.sciencedirect.com/science/article/pii/S0749641903001414}

\bibitem[{Voyiadjis and Faghihi(2013)}]{voyiadjis_2013}
Voyiadjis, G.~Z., Faghihi, D., 2013. Localization in stainless steel using
  microstructural based viscoplastic model. International Journal of Impact
  Engineering 54, 114 -- 129.
\newline\urlprefix\url{http://www.sciencedirect.com/science/article/pii/S0734743X12001868}

\bibitem[{Wright(2002)}]{wright_2002}
Wright, T.~W., 2002. The Physics and Mathematics of Adiabiatic Shear Bands.
  Cambridge University Press, Cambridge.

\bibitem[{Zaera et~al.(2013)Zaera, Rodríguez-Martínez, and
  Rittel}]{zaera_2013}
Zaera, R., Rodríguez-Martínez, J., Rittel, D., 2013. On the taylor-quinney
  coefficient in dynamically phase transforming materials. application to 304
  stainless steel. International Journal of Plasticity 40, 185 -- 201.
\newline\urlprefix\url{http://www.sciencedirect.com/science/article/pii/S0749641912001192}

\bibitem[{Zehnder(1991)}]{zehnder_1991}
Zehnder, A.~T., 1991. A model for the heating due to plastic work. Mechanics
  Research Communications 18~(1), 23 -- 28.
\newline\urlprefix\url{http://www.sciencedirect.com/science/article/pii/009364139190023P}

\bibitem[{Zhao et~al.(2016)Zhao, Low, Wang, and Niezgoda}]{zhao_2016}
Zhao, P., Low, T. S.~E., Wang, Y., Niezgoda, S.~R., 2016. An integrated
  full-field model of concurrent plastic deformation and microstructure
  evolution: Application to 3d simulation of dynamic recrystallization in
  polycrystalline copper. International Journal of Plasticity 80, 38 -- 55.
\newline\urlprefix\url{http://www.sciencedirect.com/science/article/pii/S0749641915002156}

\bibitem[{Zhao et~al.(2018)Zhao, Wang, and Niezgoda}]{zhao_2018}
Zhao, P., Wang, Y., Niezgoda, S.~R., 2018. Microstructural and micromechanical
  evolution during dynamic recrystallization. International Journal of
  Plasticity 100, 52 -- 68.
\newline\urlprefix\url{http://www.sciencedirect.com/science/article/pii/S0749641917304345}

\end{thebibliography}

%% else use the following coding to input the bibitems directly in the
%% TeX file.

%\begin{thebibliography}{00}

%% \bibitem{label}
%% Text of bibliographic item

%\bibitem{}

%\end{thebibliography}
\end{document}